\documentclass{JHEP3} 

\usepackage{graphicx}
\usepackage{subfigure}
\newcommand\fverb{\setbox\fverbbox=\hbox\bgroup\verb}
\newcommand\fverbdo{\egroup\medskip\noindent%
            \fbox{\unhbox\fverbbox}\ }
\newcommand\fverbit{\egroup\item[\fbox{\unhbox\fverbbox}]}
\newbox\fverbbox

\newcommand{\be}{\begin{equation}}
\newcommand{\ee}{\end{equation}}
\newcommand{\bea}{\begin{eqnarray}}
\newcommand{\eea}{\end{eqnarray}}
\newcommand{\ba}{\begin{array}}
\newcommand{\ea}{\end{array}}

\def\a{\alpha}
\def\b{\beta}

\def\o{\omega}

\def\r{\rho}

\def\t{\tau}

\def\O{\Omega}

\title{Holographic meson mass in asymmetric dense matter }

\author{Youngman Kim
        \\
Asia Pacific Center
for Theoretical Physics and Department of Physics, Pohang University
of Science and Technology, Pohang, Gyeongbuk 790-784, Korea
\\  E-mail: \email{ykim@apctp.org}}

\author{Yunseok Seo
\\
Center for Quantum Spacetime, Sogang University, Seoul 121-742, Korea
\\ E-mail: \email{yseo@sogang.ac.kr}}

\author{Ik Jae Shin
\\
Asia Pacific Center
for Theoretical Physics, Pohang, Gyeongbuk 790-784, Korea
\\ E-mail: \email{geniean@apctp.org}}

\author{Sang-Jin Sin
\\
Department of Physics, Hanyang University, Seoul 133-791, Korea
\\ E-mail: \email{sjsin@hanyang.ac.kr}}
\received{\today}       

\abstract{We study a meson mass splitting due to isospin violation in holographic dense matter.
We work in a D4/D6/D6 model with two quark flavor branes  to consider asymmetric dense matter in holographic QCD.
We  mainly consider two cases.
We first consider $m^+/m^-\sim m_d/m_u$ to study the effect of isospin violation on the meson masses.
Then, we  take $m^+/m^-\sim m_s/m_q$, where $m_q\sim(m_u+m_d)/2$, to calculate in-medium kaon-like meson masses.
In both cases we observe that the mass splitting of charged mesons occurs at low densities due to the asymmetry, while
 at high densities their masses become degenerate. At intermediate densities, we find an exotic behavior in masses
 which could be partly understood in a simple picture based on the Pauli exclusion principle.}


\begin{document}

\section{Introduction}
Asymmetric dense matter is a fascinating subject in modern nuclear physics.
Nature provides isospin asymmetric nuclear matter in the interior of a nucleus
with unequal number of protons and neutrons and in compact stars such as a neutron star.
Through heavy ion collisions with neutron rich, stable and/or radioactive nuclei, we could study the properties of
this matter in the laboratory.
As density goes up, strangeness starts to play important roles. For instance,
the transition from nuclear matter to strange matter, through onset of hyperon matter or kaon condensation,
shows a critical impact on  the properties of compact
stars such as the mass-radius relation and on core-collapse supernovae.

In addition to vacuum properties of dense matter, the properties of
hadrons in such  environment are of interest.
The in-medium mass of $K^-$ is one of the key quantities to
determine the onset of kaon condensation.
The in-medium modification of kaon/antikaon properties such as the mass can
be observed experimentally. One of the experimental observables is the $K^-/K^+$ production ratios.
Upcoming experiments such as the GSI Facility for Antiproton and Ion Research (FAIR)
would reveal the in-medium properties of heavy-light mesons, $D$ and $\bar D$, through
experimental signatures of their production ratio and $J/\psi$ suppression.
We refer to \cite{dM_review} for a review on the issues in dense matter.

The main goal of this paper is to study the in-medium meson masses and their mass splitting in an asymmetric dense matter using a
holographic QCD model.
Based on the AdS/CFT correspondence~\cite{Maldacena:1997re,Gubser:1998bc,Witten:1998qj},
a holographic model of QCD~\cite{hQCD}, see \cite{hQCD_review} for a review, is offering a new analytic tool to study dense matter.
In \cite{KSS2010}, a simple model for nuclear matter to strange matter transition has been proposed in a D4/D6/D6 model.
In this work the authors introduced two flavor D6 branes for light (u or d)  and intermediate (strange) quarks
respectively and compact D4 brane with $N_c$ fundamental strings attached. By considering the force-balancing condition and energy minimization, they could
study transition from a dense matter with only light quarks  to a matter with both light and intermediate quarks. Since the
background in \cite{KSS2010} is confining D4 geometry, the authors could associate this transition with nuclear to strange matter transition.
Our present study is based on this D4/D6/D6 model to gain asymmetry in dense nuclear matter. In this sense, the mass difference of two quarks
is the origin of our asymmetry. To be more realistic, our model should be supplied with a few more ingredients.
For instance, in neutron stars the charge neutrality favors neutrons rather than protons since protons should come in together with electrons.
A real neutron star contains a small fraction of protons and electrons to prevent neutrons from undergoing a weak decay into protons and electrons, though. Obviously we need to have three flavors with different quark masses. To this end we have to extend
the present model to a D4/D6/D6/D6 model.
With these cautions in mind, we will study the meson masses in asymmetric dense matter, hoping that
our study would catch some main features of the physics of asymmetric matter.
We will consider two cases with different quark mass ratios:
$m^+/m^-\sim m_d/m_u$ and  $m^+/m^-\sim m_s/m_q$, where $m_q=(m_u+m_d)/2$.
The first case is for a nuclear matter with isospin violation, and the second one is to mimic
a transition from nuclear matter to strange (hyperon) matter.

In Appendix \ref{la17A}, we use a different value of the 't Hooft coupling from the one used in the main text
to see if our results are sensitive to the choice of the coupling.

 \section{Asymmetric dense matter \label{adm}}
In this section we review  the work of \cite{KSS2010}, where a toy model, based on a D4/D6/D6 setup,
for nuclear matter to strange matter transition has been proposed.
In \cite{KSS2010}, the matter with one light and one intermediate quarks is named as strange matter without proper discussion of
the strangeness quantum number. This is basic reason why the authors of \cite{KSS2010}  considered their model as  a {\it toy}.
Note, however, that in some cases describing strangeness physics with two flavors works well.
For instance, the essential physics of kaon condensation could be captured in the V-spin approach, where  chiral
SU(3)$\times$SU(3) is effectively reduced to SU(2)$\times$ SU(2)~\cite{BKR}.
In this work we will make use of the asymmetry obtained in \cite{KSS2010} to study meson mass in asymmetric dense matter.

Now we review the paper \cite{Seo:2008qc, KSS2010}.
The non-supersymmetric geometry for confining background of D4 is given by
\begin{eqnarray}
ds^{2}
&=&\left(\frac{U }{R }\right)^{3/2}\left(\eta_{\mu\nu}dx^\mu dx^\nu + f(U) dx_4^{2} \right)
+\left(\frac{R}{U }\right)^{3/2}\left( \frac{dU^2}{ f(U)} +U^2 d\Omega_4^2\right) \cr
e^\phi&=&g_s\left(\frac{U }{R }\right)^{3/4},\quad F_4 =\frac{2\pi N_c}{\Omega_4}\epsilon_4, \;\; f(U)=
1-\Big(\frac{U_{KK}}{U}\Big)^{3}, \;\; R^3=\pi g_s N_c l_s^3.
\label{adsm}
\end{eqnarray}
The Kaluza-Klein mass is defined as inverse radius of the $x_4$ direction:
$M_{KK}=\frac{3}{2}\frac{U^{1/2}_{KK}}{R^{3/2}}$.
The bulk parameters, $U_{KK}, g_s$, and $R$, and the parameters of the gauge theory,
$M_{KK}$ and $g^2_{YM}$, are related by
\be\label{consts1}
g_s=\frac{\lambda}{2\pi l_sN_c M_{KK}}, \quad U_{KK}=\frac{2}{9}\lambda M_{KK} l_s^2,
\quad R^3=\frac{\lambda l_s^2}{2M_{KK}} ,\quad \lambda=g_{YM}^{2}N_{c}.
\ee
To make transverse space to D4 brane be flat (up to overall factor), we introduce new coordinate $\bar{\xi}$,  $\frac{d\bar{\xi}^2}{\bar{\xi}^2}=\frac{dU^2}{U^2f(U)}$,
and obtain, in Euclidean signature,
\be\label{d4bgmetric}
ds^2 = \left(\frac{U }{R }\right)^{3/2}\left(dt^2 +d\vec{x}^2 + f(U) dx_4^{2} \right)
+\left(\frac{R}{U }\right)^{3/2}\left(\frac{U}{\bar{\xi}}\right)^2\left(d\bar{\xi}^2 +\bar{\xi}^2 d\Omega_4^2\right)\, .
\ee
The relation between $U$ and $\xi$ is
\be \label{uxi}
U^{3/2}=\bar{\xi}^{3/2}\left[1+\left(\frac{\bar{\xi}_{KK}}{\bar{\xi}}\right)^{3}\right]\equiv\bar{\xi}^{3/2}\omega_{+} = \bar{\xi}_{KK}^{3/2} \xi^{3/2}\omega_{+},
\ee
where $U_{KK}^{3/2} =2 \bar{\xi}_{KK}^{3/2}$ and we rescale $\bar{\xi}$ coordinate such that singularity is located at $\xi=1$.
With these new coordinates we rewrite the background (\ref{d4bgmetric}) and the dilaton as
\bea\label{scaledmet}
ds^2 &=&\left(\frac{\bar{\xi}_{KK}}{R}\right)^{3/2} \xi^{3/2} \omega_{+}\left(dt^2 +d\vec{x}^2+f(U)dx_4^2\right) +\left(R^3 \bar{\xi}_{KK}\right)^{1/2} \frac{\omega_{+}^{1/3}}{{\xi}^{3/2}}\left(d{\xi}^2 +{\xi}^2 d\Omega_{4}^2\right),\cr
 &&~~~~~~~~~~~~~~~~~~~~~~~~~e^{\phi}=\left(\frac{\bar{\xi}_{KK}}{R}\right)^{3/4}{\xi}^{3/4}\omega_{+}^{1/2}.
\eea

A baryon in four-dimensional theory
could be described by a D4 brane wrapping  $S^4$.  In this configuration, the background four-form field strength
couples to the world-volume gauge field $A_{(1)}$ via the Chern-Simons interaction. The source of these world-volume gauge field can be interpreted as the
 end point of $N_c$ fundamental strings.\par
We rewrite again the background metric (\ref{scaledmet})  as
\bea\label{d4_ind}
ds^2&\!=\!&\left(\frac{\bar{\xi}_{KK}}{R}\right)^{3/2} {\xi}^{3/2} \omega_{+}\left(dt^2 +d\vec{x}^2+f(U)dx_4^2\right) \nonumber \\
&&~~~~~~~~~~+\left(R^3 \bar{\xi}_{KK}\right)^{1/2} \omega_{+}^{1/3}{\xi}^{1/2}\left(\frac{d{\xi}^2}{{\xi}^2} +d\theta^2+\sin^2\theta d\Omega_{3}^2\right).
\eea
We take $(t,\theta,\theta_\a)$ as a world-volume coordinate of D4 brane,
where $\theta_\a$ are angular coordinates on $S^3$ and turn on the $U(1)$ gauge field on it. For simplicity, we assume that the position of D4 brane and the gauge field depends only on $\theta$, {\it i.e.},
${\xi}={\xi}(\theta)$, $A_{t}=A_t(\theta)$, where $\theta$ is the polar angle.
The induced metric on the compact D4 brane is
\be\label{d4met}
ds^2 =\left(\frac{\bar{\xi}_{KK}}{R}\right)^{3/2} {\xi}^{3/2} \omega_{+} dt^2+\left(R^3 \bar{\xi}_{KK}\right)^{1/2} \omega_{+}^{1/3}{\xi}^{1/2}\left[\left(1+\frac{{\xi^\prime}^2}{\xi^2}\right)d\theta^2+\sin^2\theta d\Omega_{3}^2\right].
\ee
Then the DBI action for the single D4 brane reads
\bea\label{bary-d4}
S_{D4} &=& -\mu_4 \int e^{-\phi} \sqrt{{\rm det}(g+2\pi \alpha' F)}+\mu_4 \int  A_{(1)}\wedge G_{(4)} \cr\cr
&=& \t_4 \int dtd\theta \sin^3\theta
\left[-\sqrt{ \o_+^{4/3} (\xi^2 +\xi'^2)-\tilde{F}^2}
+3 \tilde{A}_t \right] \cr
&=& \int dt {\cal L}_{D4},
\eea
where
\be
\tau_4 = \mu_4 \O_3 g_s^{-1} R^{3}\bar{\xi}_{KK},~~~~~~\tilde{A}_t=\frac{ 2\pi\a'A_t}{\bar{\xi}_{KK}}.
\ee
Substituting the solution of equation of motion for gauge field $\tilde{A}_t$, we get `Hamiltonian' of D4 brane as
\bea\label{d4h}
{\cal H}_{D4} &=&\tilde{F}\frac{\partial {\cal L}_{D4}}{\partial \tilde{F}}-{\cal L}_{D4}\cr\cr
&=&\t_4 \int d\theta \sqrt{\omega_+^{4/3} (\xi^2 +\xi'^2)}\sqrt{D(\theta)^2+ \sin^6\theta},
\eea
where
\be\label{displacement}
D(\theta)=-2+3(\cos\theta -\frac{1}{3}\cos^3\theta).
\ee
Here, we set all fundamental strings are attached on north pole of D4 brane, because we are interested in the system in which fundamental strings connect baryon D4 brane and probe D6 branes as discussed in \cite{callan, Seo:2008qc}. \par
After  imposing boundary
condition $\xi'(0)=0$ and $\xi(0)=\xi_0$ at $\theta=0$, we can get numerical solutions which are parameterized by initial value of $\xi_0$.
The solutions corresponding different $\xi_0$'s
are drawn in Fig. 1(a).

\begin{figure}[!ht]\label{fig:baryon-d4}
\begin{center}
\subfigure[]{\includegraphics[angle=0, width=0.3\textwidth]{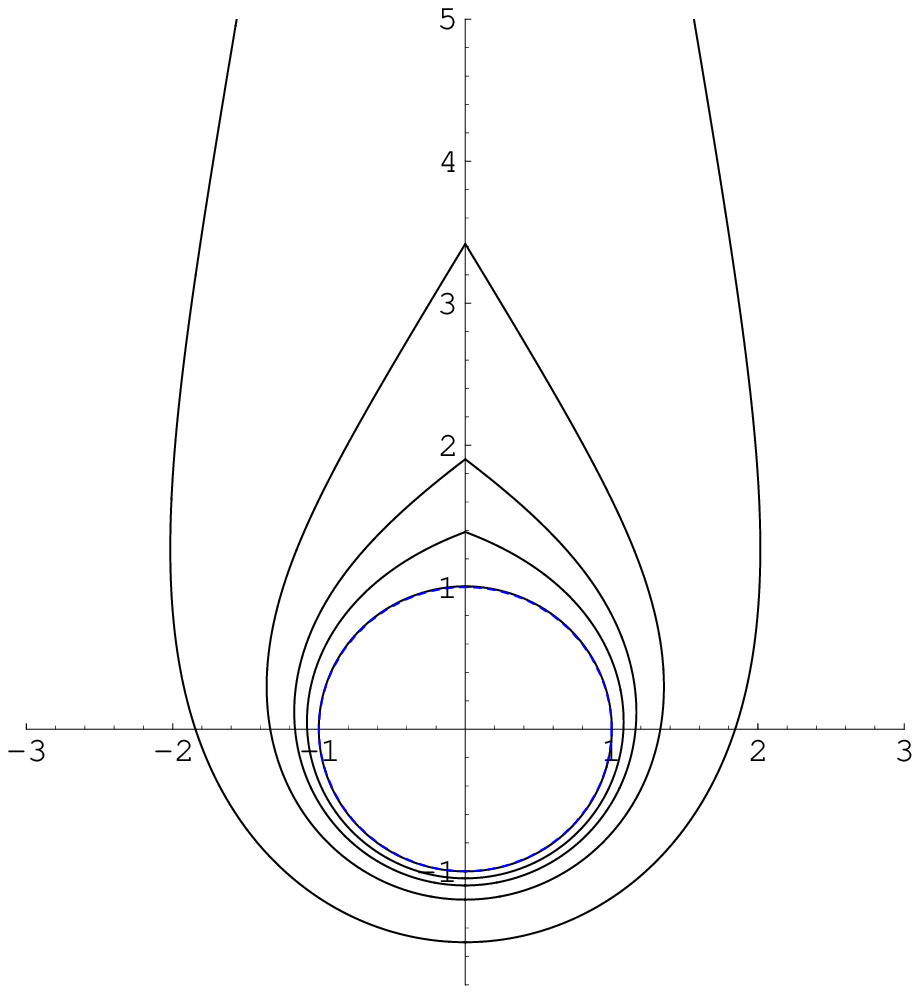}}
~~~~~
\subfigure[]{\includegraphics[angle=0, width=0.5\textwidth]{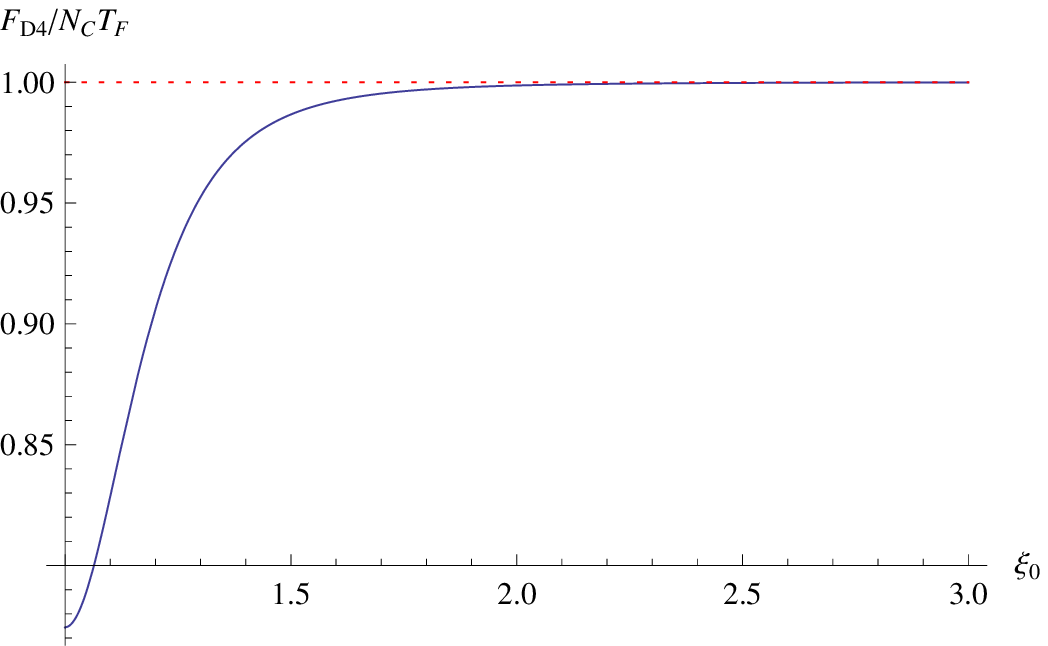}}
\caption{(a) Shape of D4 brane in $(\xi,~\theta)$ plane for different $\xi_0 = 1.001,~1.05,~1.1,~1.2,~1.5$ from inside to outside. Here south pole corresponds to $\theta=0$. (b) Force at the cusp of D4 brane for different $\xi_0$.}
\end{center}
\end{figure}

If we denote the position of the cusp of D4 brane  by $U_c$, the force at the cusp due to the D4 brane tension
can be obtained by  varying the Hamiltonian of D4 brane with respect to $U_c$ while keeping other variables fixed;
\bea\label{force-d4}
F_{D4}&=& \frac{\partial{{\cal H}}}{\partial U_c} \Bigg|_{\rm fix~other~values} \cr\cr
&=& N_c T_F \left(\frac{1+\xi_c^{-3}}{1-\xi_c^{-3}}\right)
\frac{\xi_c'}{\sqrt{\xi_c^2 +\xi_c'^2}},
\eea
where $T_F =\frac{2^{2/3}\t_4 }{N_c U_{KK}}$ is the tension of the fundamental string.
 As shown in Fig. 1(b),
the tension at the cusp of D4 brane is always smaller than the tension of the $N_c$ fundamental strings.
Therefore, if there are no other objects, the cusp should be pulled up to infinity  and the final configuration of D4
brane would be `tube-like' shape as in  \cite{callan}. To have a stable configuration,
we put probe D6 branes which the other end point of the fundamental string can be attached.

Now we consider the system with two flavors, one light and one intermediate mass quarks, by introducing two D6 branes.
To study dynamics of multi D-branes, we start from the non-Abelian DBI action \cite{naDBI}.
 \begin{equation}\label{nonAbelianDBI}	
 S\!=\!-\mu_6\!\int\!\!d^7\!\sigma~\textrm{STr}\Big[e^{-\Phi}\!\sqrt{-\textrm{det}
 \big(P[G_{rs}+G_{ra}(Q^{-1}\!-\delta)^{ab}G_{bs}]+T^{-1}F_{rs}\big)}\sqrt{\textrm{det}\,Q^a_{\;b}}\Big]\, ,
 \end{equation}
 where  STr denotes the symmetrized trace for flavor indices and P[~] does the pull-back of $10$D tensor to the world-volume of the branes.
 The matrix $Q^a_{\;b}$ is defined as
 \begin{equation}
	Q^a_{\;b}\equiv\delta^a_{\;b}+iT[X^a,X^c]G_{cb}\, ,
 \end{equation}
 where $T\!=\!1/(2\pi\alpha^\prime)$ and $X^a$'s are the coordinates being transverse to the branes which are $2\times 2$ matrix valued functions in general.

To obtain the embedding solution of two D6 branes, we rewrite the bulk metric (\ref{scaledmet}) as
\bea
ds^2&\!=\!&\left(\frac{\bar{\xi}_{KK} }{R }\right)^{3/2}\xi^{3/2}\omega_{+}\left(dt^2 +d\vec{x}^2 + f(U) dx_4^{2} \right) \nonumber \\
&&~~~~~~~~~~+\left(R^3 \bar{\xi}_{KK}\right)^{1/2}\frac{\omega_{+}^{1/3}}{\xi^{3/2}}\left(d\r^2 +\r^2 d\Omega_2^2+dy^2 +y^2 d\phi^2\right),
\eea
where D6 brane world-volume coordinates are $(t,\vec{x},\r,\theta_{\a})$.
We assume that the transverse coordinates depend only on $\r$.
The induced metric on D6 brane is
\bea
ds^2&\!=&\!\left(\frac{\bar{\xi}_{KK} }{R }\right)^{3/2}\xi^{3/2}\omega_{+}\left(dt^2 +d\vec{x}^2  \right) \cr
&&~~~~~~~~~~+\left(R^3 \bar{\xi}_{KK}\right)^{1/2}\frac{\omega_{+}^{1/3}}{\xi^{3/2}}\left[\left(1+y'^2 +y^2 \phi'^2\right)d\r^2 +\r^2 d\Omega_2^2\right],
\eea
where prime denotes to derivative with respect to $\r$. Here, $y$ and $\phi$ which transverse to D6 brane world-volume are matrix-valued functions.  \par
In this work, we take the diagonal ansatz for embedding functions, which is physically relevant since we expect off-diagonal quark condensates to be zero, $<\bar u d>=<\bar du>=0$.
\bea
y(\rho)=\left(
		\begin{array}{cc}
			\!\!y^{+}(\rho)\!~&~\!\!0\!\! \\
			\!\!0\!~&~\!\!y^{-}(\r)\!\!
		\end{array}\right),
~~~~~~
\phi =0,
\eea
where $y^{\pm}$ is the position of each probe brane.
On each brane, we turn on $U(1)$ gauge field whose source is the end point of fundamental strings
\bea
A_t(\rho)=\left(
		\begin{array}{cc}
			\!\!A_t^{+}(\rho)\!~&~\!\!0\!\! \\
			\!\!0\!~&~\!\!A_t^{-}(\rho)\!\!
		\end{array}\right).
\eea
In terms of the Pauli matrix
 \begin{equation}
	\tau^0=\frac{1}{2}\;\mathbb{I}_{2\times2}~,~~\tau^1=\frac{1}{2}\left(
		\begin{array}{cc}
			\!\!0\!~&~\!\!1\!\! \\
			\!\!1\!~&~\!\!0\!\!
		\end{array}
	\right)~,~~\tau^2=\frac{1}{2}\left(
		\begin{array}{cc}
			\!\!0\!&~\!\!-i\!\! \\
			\!\!i\!&~\!\!0\!\!
		\end{array}
	\right)~,~~\tau^3=\frac{1}{2}\left(
		\begin{array}{cc}
			\!\!1\!&~\!\!0\!\! \\
			\!\!0\!&~\!\!-1\!\!
		\end{array}
	\right)\;,
 \end{equation}
$y$ can be written as
\be
y = y^0 \tau^0 + y^3 \tau^3,
\ee
where
\be
y^0 = y^+ +y^-, ~~~~~~y^3 =y^+ -y^-.
\ee
Near the boundary, $y^{0,3}$ becomes
\bea
y^0 &\sim& \left(m^+ +m^-\right) +\frac{c^+ +c^-}{\r^2} + \cdots\;, \cr
y^3 &\sim& \left(m^+ -m^-\right) +\frac{c^+ -c^-}{\r^2} + \cdots\;.
\eea
If we put two branes on top of each other, the value of $y^3$ becomes zero at the boundary.
This is physically obvious since the exact two flavor isospin symmetry U(2) guarantees
$m^+ = m^-$ and $c^+ = c^-$.

Similarly, $U(1)$ gauge field can be written as
\be
A_t = A^0 \tau^0 + A^3 \tau^3.
\ee
Near boundary, we can define isospin chemical potential and isospin charge as
\bea\label{isospin}
A^0 &=& (\mu^+ + \mu^-) + \frac{Q^+ +Q^-}{\r^2} + \cdots \equiv \mu_T + \frac{Q_T}{\r^2} + \cdots\;,\cr
A^3 &=&  (\mu^+ - \mu^-) + \frac{Q^+ -Q^-}{\r^2} + \cdots \equiv \mu_I + \frac{Q_I}{\r^2} + \cdots\;.
\eea

\begin{figure}[!ht]
\begin{center}
\subfigure[]{\includegraphics[angle=0, width=0.45\textwidth]{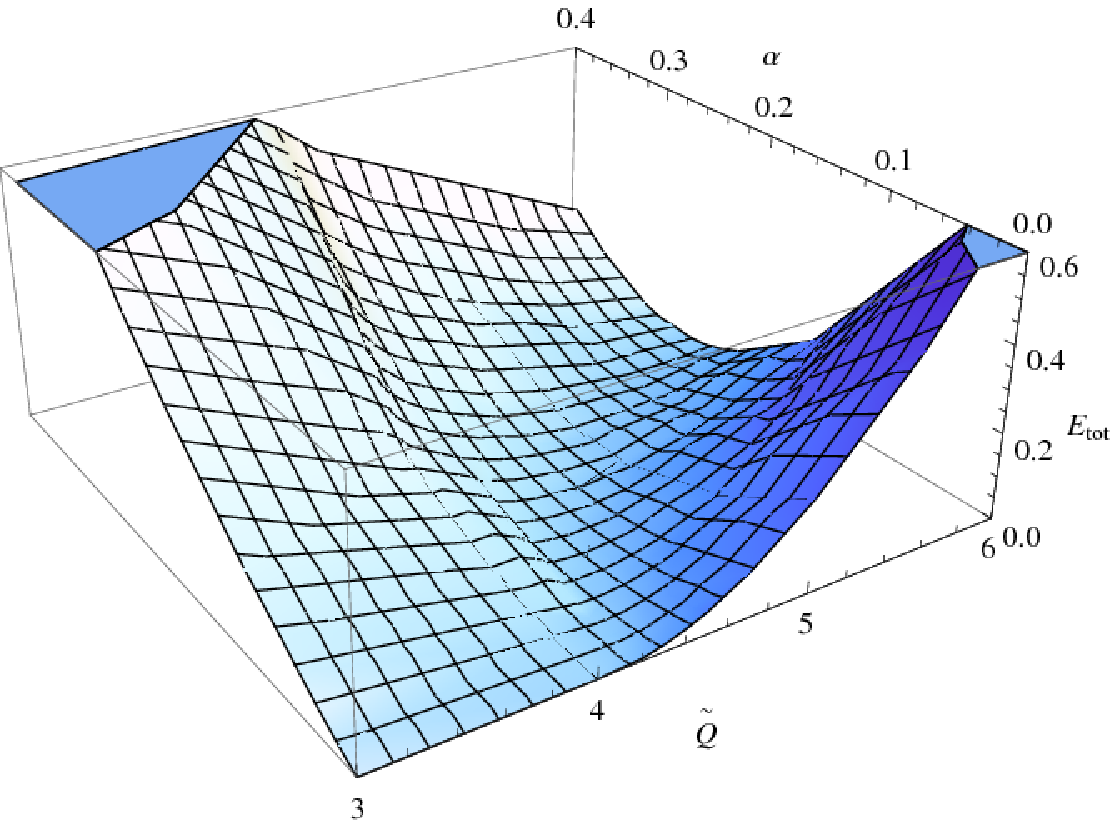}}
~~~~
\subfigure[]{\includegraphics[angle=0, width=0.5\textwidth]{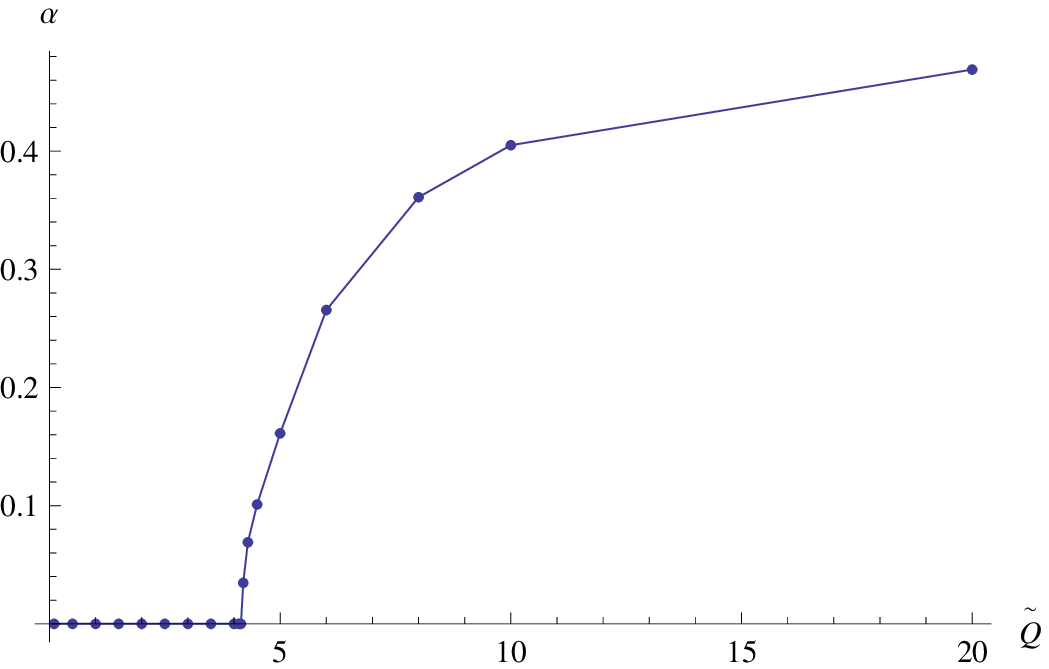}}
\caption{
(a) $\a$ dependence of regularized free energy.\label{fig:faq}
(b) Density dependence of $\a$.
}
\end{center}
\end{figure}

 With this diagonal ansatz, the commutators in the non-Abelian DBI action (\ref{nonAbelianDBI}) vanish, and the action can be expressed as the sum of two independent embeddings
\bea\label{d6DBI}
S_{D6}&=&  -\t_6 \int dtd\r\,\r^2  \o_+^{4/3} \sqrt{\o_+^{4/3} (1+y'^2)-\tilde{F}^2}\Bigg|_{+}\cr
&&~~~~~-\t_6 \int dtd\r\,\r^2  \o_+^{4/3} \sqrt{\o_+^{4/3} (1+y'^2)-\tilde{F}^2}\Bigg|_{-},
\eea
where
\bea\label{consts2}
\t_6 = \mu_6 V_3  \O_2 g_s^{-1} \bar{\xi}_{KK}^3 ,\quad
\tilde{F}=\frac{2\pi\a'F_{t\r}}{\bar{\xi}_{KK}}\, .
\eea
Here $\pm$ denotes that each DBI action is function of $y^{\pm}$ and $A_t^{\pm}$ separately.
We define a dimensionless quantity $\tilde{Q}^{\pm}$ from the equation of motion for $\tilde{F}^{\pm}$;
\be
\frac{\partial S_{D6}}{\partial \tilde{F}^{\pm}}
=\frac{ \r^2 \o_+^{4/3}\tilde{F}}{\sqrt{\o_+^{4/3} (1+y'^{2})-\tilde{F}^2}}\Bigg|_{\pm}
\equiv \tilde{Q}^{\pm}.
\ee
It is related to the number of point sources (number of fundamental strings) $Q^{\pm}$  by
\be
\tilde{Q}^{\pm}=\frac{\bar{\xi}_{KK}Q^{\pm}}{2\pi\a'\t_6}.
\ee
Through the Legendre transformation, we obtain the Hamiltonian
\bea\label{d6h}
{\cal H}_{D6} &=&\tilde{F}\frac{\partial S_{D6}}{\partial \tilde{F}}-S_{D6} \cr\cr
&=& \t_6 \int d\r \sqrt{\o_+ (y^{+})^{4/3}\left(\tilde{Q}^{+2}+\r^4 \o_+ (y^{+})^{8/3}\right)}\sqrt{1+{y^+}'^2
} \cr
&&+\t_6 \int d\r \sqrt{\o_+ (y^{-})^{4/3}\left(\tilde{Q}^{-2}+\r^4 \o_+ (y^{-})^{8/3}\right)}\sqrt{1+{y^-}'^2
}\cr\cr
&=& {\cal H}_{D6}(\tilde{Q}^+,y^{+};\r) +{\cal H}_{D6}(\tilde{Q}^-,y^{-};\r).
\eea
Since the Hamiltonian of two D6 branes is split into two independent Hamiltonian of each brane, we can get the embedding solution of each brane separately. In our configuration, $Q^+$ fundamental strings are attached on the upper brane and $Q^-$ strings end on the lower brane, while the other end points of the fundamental strings are attached on the D4 brane.
Therefore, we expect that two probe branes and spherical D4 brane meet at the same point $\xi_c$ eventually.
The force at the cusp of the D6 branes is given by
\bea\label{force-d6}
F_{D6}&=&\frac{\partial {\cal H}_{D6}(Q^+)}{\partial U_c} \Bigg|_{\partial}
+\frac{\partial {\cal H}_{D6}(Q^-)}{\partial U_c} \Bigg|_{\partial} \cr\cr
&=&\frac{Q^+}{2\pi\a'}\left(\frac{1+\xi_c^{-3}}{1-\xi_c^{-3}}\right)
\frac{{y_c^{+}}'}{\sqrt{1+{y_c^{+}}'^{2}}}
+\frac{Q^-}{2\pi\a'}\left(\frac{1+\xi_c^{-3}}{1-\xi_c^{-3}}\right)
\frac{{y_c^{-}}'}{\sqrt{1+{y_c^{-}}'^{2}}}\cr\cr
&\equiv& F_{D6}^{+}(Q^+)+F_{D6}^{-}(Q^-),
\eea
where ${y_c^{\pm}}'$ denotes the slope of each brane at the cusp.
To make the whole system stable, the force at the cusp of D4 brane should be balanced to those of D6 branes;
\be\label{fbc}
\frac{Q}{N_c} F_{D4} =  F_{D6}^{+}(Q^+)+F_{D6}^{-}(Q^-),
\ee
where $Q=Q^+ +Q^-$.
Rewriting $Q^{\pm} $ and ${y_c^{\pm}}'$ by using new parameters $\a$ and $\b$,
\bea
Q^{+} =\a Q~~&,& Q^{-} =(1-\a)Q, \cr
{y_c^{+}}'=\b {y_c}'&,&~~~ {y_c^{-}}'={y_c}',
\eea
the force balance condition (\ref{fbc}) becomes
\be\label{fbc2}
\frac{\xi'_c}{\sqrt{\xi_c^{'2} +\xi_c^{2}}} =\frac{\a\b{y_c}'}{\sqrt{1+\b^2{y_c}'^{2}}}
+\frac{(1-\a){y_c}'}{\sqrt{1+{y_c}'^{2}}}.
\ee
Together with the force balancing condition, we consider the energy minimization of the system.
The total energy of the system is
\be \label{Etot}
E_{tot}=\frac{Q}{N_c} {\cal H}_{D4} +{\cal H}_{D6}(Q^+) +{\cal H}_{D6}(Q^-)
=\tau_6 \left[ \frac{\tilde{Q}}{4} E_4 +E_6 (\tilde{Q}^+)+E_6 (\tilde{Q}^-)\right],
\ee
where $E_i$ is numerical integration of each `Hamiltonian' in (\ref{d4h}) and  (\ref{d6h}) without overall constant
$\tau_4$ and $\tau_6$.
As an example, we fix the asymptotic values of probe branes as $m^+ =5$ and $m^-=0.1$.
The $\alpha$ dependence of the total energy with different densities is drawn in Fig. \ref{fig:faq}(a)
 and the density dependence of $\alpha$ for stable and minimum-energy configuration is shown in Fig. \ref{fig:faq}(b).
 In low density, only the lower brane ($m^- =0.1$) constitutes a physical configuration.
 As density increases, at certain density, baryon vertex which attached to the upper brane can come into the system.
And, at high densities the system consists of baryons attached to upper and lower branes equally.

\begin{figure}[!ht]
\begin{center}
\subfigure[]{\includegraphics[angle=0, width=0.4\textwidth]{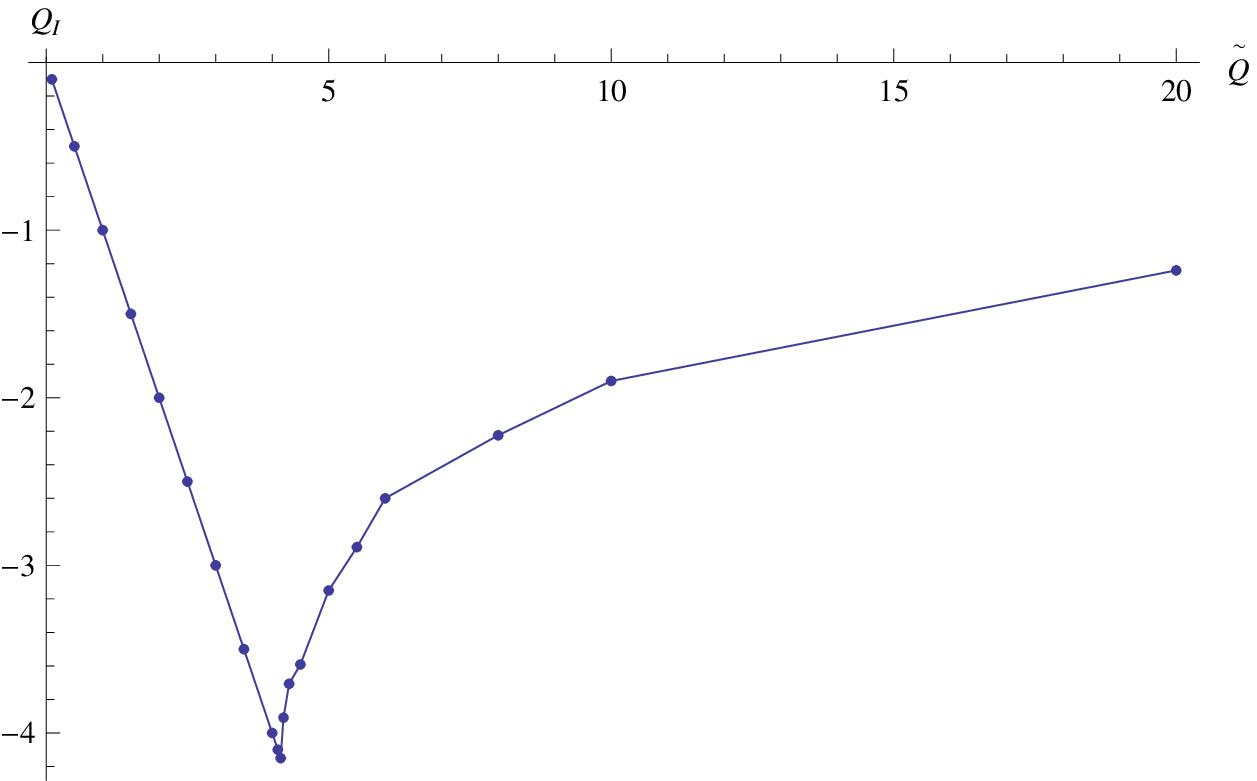}}
~~~
\subfigure[]{\includegraphics[angle=0, width=0.4\textwidth]{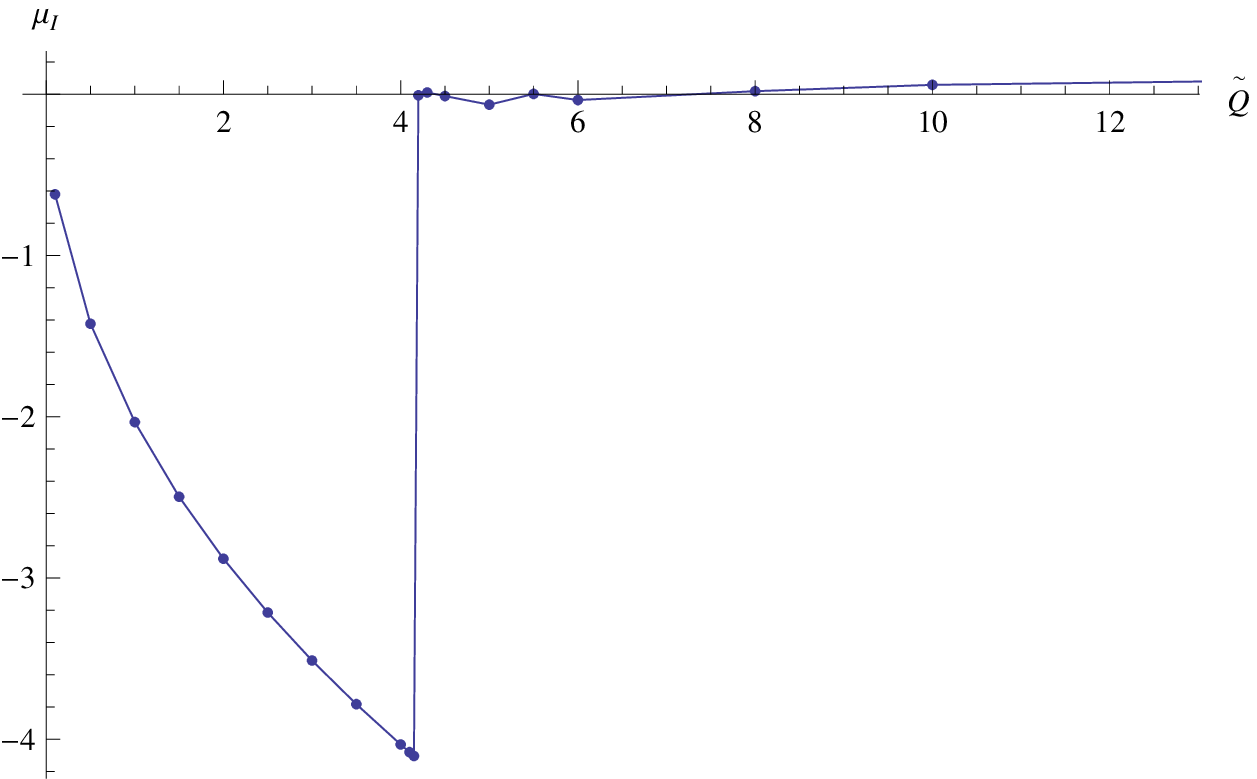}}

\caption{(a) Density dependence of the density difference. (b) Density dependence of the chemical potential difference.\label{fig:muI}}
\end{center}
\end{figure}

We can also calculate the chemical potential difference of two quarks, $m^+$ and $m^-$, and its canonical conjugate density by using (\ref{isospin}). Numerical results of density dependence of these quantities are drawn in Fig. \ref{fig:muI}.
 At low density, all fundamental strings are attached on the lower brane, i.e. $Q^+ =0$. In this regime,
 the density difference is nothing but $Q_I = -Q^-$, and it decreases linearly up to the transition density. Above the transition point, the number of fundamental strings on the upper brane increases so that the difference between $Q^+$ and $Q^-$ decreases.
 Therefore, the density difference increases and it approaches zero at large density as in Fig. \ref{fig:muI}(a).

The density dependence of the chemical potential difference is shown in Fig. \ref{fig:muI}(b).
The behavior at low density is similar to the density difference.
However, at the transition density, the chemical potential difference suddenly becomes zero and keeps this value for all density afterwards.
This behavior is quite reasonable in the following sense. A simple reason for the transition from $\alpha=0$ to $\alpha\neq 0$
is that the chemical potential (or Fermi energy) of light quark $\mu^-$ becomes comparable to the mass of heavier one $m^+$, $\mu^-\sim m^+$,
near the transition density.
While, just above the transition point,  the chemical potential of the heavier quark $\mu^+$ will be close to its mass, $\mu^+\sim m^+$.
Therefore we could expect $\mu^+-\mu^-\sim 0$ at and just above the transition density.
As density increases further, the chemical potential difference will be stay close to zero since
at high densities the quark mass difference could be neglected.

\section{Meson mass in asymmetric matter}
We begin this section with a simple example of isospin matter to gain a rough picture on the
behavior of meson masses in asymmetric matter.
Isospin matter with finite isospin chemical potential
and zero baryon chemical potential was proposed as a useful system to improve our
understanding of cold dense QCD~\cite{SSiso}, though this kind of matter hardly exists in  nature.
At small (negative) isospin chemical potential $\mu_I$ with zero isospin number density, the mass of the pion is given by
\bea
m_{\pi^\pm}=m_{\pi^0} +q|\mu_I|\, ,
\eea
where $q$ is the isospin charge of the particle.
Note that $\pi^0$ has zero isospin charge and so the mass will be intact with a finite isospin chemical potential.
Since $\pi^-$ and $\pi^+$ have the opposite isospin charge $q=-1$ and $q=+1$ respectively, their mass will be splitting with increasing
$|\mu_I|$, linearly in this case.
At some large value of the chemical potential $|\mu_I|$, pion condensation will set in, and the in-medium behavior of the pion mass becomes more intriguing.
For more details on isospin matter we refer to~\cite{SSiso}.

Now we calculate meson masses in asymmetric matter based on \cite{KSS2010}.
To this end, we consider non-Abelian DBI proposed in \cite{naDBI}.
We will basically consider two cases.
We first consider $m^+/m^-\sim m_d/m_u$ to study the effect of isospin violation on the meson masses.
Then we will take $m^+/m^-\sim m_s/m_q$, where $m_q=(m_u+m_d)/2$, to calculate in-medium kaon-like meson masses.

We remark that  we are focusing on a general tendency of the in-medium meson mass not on the exact numbers since our model is yet not that close to
a real QCD.

 \subsection{D$6$ brane fluctuations}
 We start from the non-Abelian DBI action (\ref{nonAbelianDBI}) for D$6$-branes in the confining D4 background.
Since we are considering one light and one intermediate mass quarks, we take $N_f=2$.
 \begin{equation}	 S\!=\!-\mu_6\!\int\!\!d^7\!\sigma~\textrm{STr}\Big[e^{-\Phi}\!\sqrt{-\textrm{det}
 \big(P[G_{rs}+G_{ra}(Q^{-1}\!-\delta)^{ab}G_{bs}]+T^{-1}F_{rs}\big)}\sqrt{\textrm{det}\,Q^a_{\;b}}\Big]
 \end{equation}
 For later convenience, we define some notations on the flavor matrix.
 In our case with two flavors, these matrices can be represented by $2\times2$-matrices
 \begin{eqnarray}
 	M=M_0\tau^0+M_i\tau^i=\frac{1}{2}
 	\left(
		\begin{array}{cc}
			\!\!M_0+M_3\!~&~\!\!M_1-iM_2\!\! \\
			\!\!M_1+iM_2\!~&~\!\!M_0-M_3\!\!
		\end{array}
	\right)\equiv\left(
		\begin{array}{cc}
			\!\!M_+\!~&~\!\!M_{12}\!\! \\
			\!\!M_{21}\!~&~\!\!M_-\!\!
		\end{array}
	\right)\, ,
 \end{eqnarray}
 using Pauli matrices.
 For diagonal brane embeddings the above action is expanded as
 \begin{eqnarray}
 S&=&-\mu_6\!\int\!\!d^7\sigma~\textrm{STr}\Big[e^{-\Phi}\sqrt{-\textrm{det}\big\{G_{rs}+D_{r}X^{a}D_{s}X^{b}
 \big(G_{ab}-iT[X^c,X^d]G_{ac}G_{db}\big)+T^{-1}F_{rs}\big\}} \nonumber \\
	&&~~~~~~~~~~~~~~~~~~~~~~~~~~~~~~~~~~~~~~~~~~~~~\times\big(1+\frac{\;T^2}{4}[X^a,X^b]G_{bc}[X^c,X^d]G_{da}\big)\Big]\;,\label{diaS}
 \end{eqnarray}
 where the covariant derivative is defined by
 \begin{equation}
	D_sX^a\equiv\partial_sX^a+i[A_s,X^a]\;.
 \end{equation}
 The induced metric on the D6 brane takes the form
 \be
ds^2= \left(\frac{\bar{\xi}_{KK} }{R }\right)^{3/2}\xi^{3/2}\omega_{+}\left(-dt^2 +d\vec{x}^2  \right)
+\left(R^3 \bar{\xi}_{KK}\right)^{1/2}\frac{\omega_{+}^{1/3}}{\xi^{3/2}}\left[\left(1+{y'}^2\right)d\r^2 +\r^2 d\Omega_2^2\right],
\ee
where $y'=\partial y/\partial\rho$.
 Embedding solutions for $y$ are summarized in Sec. \ref{adm}.

Now we consider some fluctuations to study meson spectrum in asymmetric dense matter.
\begin{equation}
	X^8=\bar{X}^8+\varphi^8=y+\varphi^8~~~,~~X^9=\bar{X}^9+\varphi^9=0+\varphi^9~~~\textrm{and}~~A_s=\bar{A}_s+\alpha_s=\delta_{s(t)}A+\alpha_s
\end{equation}
 Since the only non-vanishing background gauge field is $A_t=A$, as in Sec. \ref{adm},
 we can put $\bar{A}_s=\delta_{s(t)}A$, where $(t)$ denotes the time component.
 Then the covariant derivative is expressed as
 \begin{eqnarray}
	D_sX^a\!\!&\equiv&\!\!\partial_sX^a+i[A_s,X^a] \nonumber \\
	&=&\!\!\partial_s\bar{X}^a+\partial_s\varphi^a+i[\bar{A}_s,\varphi^a]+i[\alpha_s,\bar{X}^a]+i[\alpha_s,\varphi^a]\;.
 \end{eqnarray}
 Now we expand the action (\ref{diaS}) in terms of these fluctuations $\varphi^a$.
 To evaluate the determinant in the action, we introduce $a_{rs}$
 \begin{eqnarray}
	a_{rs}&\equiv&G_{rs}+D_{r}X^{a}D_{s}X^{b}\big(G_{ab}-iT[X^c,X^d]G_{ac}G_{db}\big)+T^{-1}F_{rs} \nonumber \\
	&=&a^{(0)}_{rs}+a^{(1)}_{rs}+a^{(2)}_{rs}+\cdots\, ,
 \end{eqnarray}
 where
 \begin{eqnarray} \label{pertDet}
	a^{(0)}_{rs}&\!=\!&G_{rs}+G_{yy}\partial_r y\partial_s y+T^{-1}(\partial_r\bar{A}_s-\partial_s\bar{A}_r)\;, \nonumber \\
	a^{(1)}_{rs}&\!=\!&G_{yy}\big\{\partial_r y(\partial_s\varphi^8+i[\bar{A}_s,\varphi^8]+i[\alpha_s,y])+(\partial_r\varphi^8+i[\bar{A}_r,\varphi^8]+i[\alpha_r,y])\partial_s y\big\} \nonumber \\
	&&~+\,T^{-1}(\partial_r\alpha_s-\partial_s\alpha_r+i[\bar{A}_r,\alpha_s]+i[\alpha_r,\bar{A}_s])\;, \nonumber \\
	a^{(2)}_{rs}&\!=\!&G_{yy}\big\{i\partial_r y[\alpha_s,\varphi^8]+i[\alpha_r,\varphi^8]\partial_s y+(\partial_r\varphi^8+i[\bar{A}_r,\varphi^8]+i[\alpha_r,y])(\partial_s\varphi^8+i[\bar{A}_s,\varphi^8]+i[\alpha_s,y])\big\} \nonumber \\
	&&~+G_{\phi\phi}(\partial_r\varphi^9+i[\bar{A}_r,\varphi^9])(\partial_s\varphi^9+i[\bar{A}_s,\varphi^9]) \nonumber \\
	&&~-iTG_{yy}G_{\phi\phi}\big\{\partial_r y(\partial_s\varphi^9+i[\bar{A}_s,\varphi^9])-(\partial_r\varphi^9+i[\bar{A}_r,\varphi^9])\partial_s y\big\}[y,\varphi] \nonumber \\
	&&~+iT^{-1}[\alpha_r,\alpha_s]\;.
 \end{eqnarray}
 Then we arrive at
 \begin{eqnarray}
	\sqrt{-\textrm{det}(a_{rs})}&\!=\!&\!\sqrt{-\textrm{det}(a^{(0)})\cdot\textrm{det}[1+(a^{(0)})^{-1}(a^{(1)}+a^{(2)})]} \nonumber \\
	 &\!=\!&\!\sqrt{-\textrm{det}(a^{(0)})}\Big(1+\frac{1}{2}\textrm{tr}[(a^{(0)})^{-1}a^{(1)}]+\frac{1}{8}\big(\textrm{tr}[(a^{(0)})^{-1}a^{(1)}]\big)^2 \nonumber \\
	&&~~~~~~~~~~~~~~~~~~-\frac{1}{4}\textrm{tr}[\big((a^{(0)})^{-1}a^{(1)}\big)^2]+\frac{1}{2}\textrm{tr}[(a^{(0)})^{-1}a^{(2)}]+\cdots\Big)\;.
 \end{eqnarray}
 Finally, we obtain the action expanded up to second-order in the fluctuation.
 \begin{eqnarray}
 S\!\!&=&\!-\mu_6\!\int\!\!d^7\sigma~\textrm{STr}\Big[e^{-\Phi}\!\sqrt{-\textrm{det}(a^{(0)})}\Big(1+\frac{1}{2}\textrm{tr}[(a^{(0)})^{-1}a^{(1)}]+\frac{1}{8}\big(\textrm{tr}[(a^{(0)})^{-1}a^{(1)}]\big)^2 \nonumber \\
 &&~~~~~-\frac{1}{4}\textrm{tr}[\big((a^{(0)})^{-1}a^{(1)}\big)^2]+\frac{1}{2}\textrm{tr}[(a^{(0)})^{-1}a^{(2)}]+\frac{\;T^2}{4}[X^a,X^b]G_{bc}[X^c,X^d]G_{da}+\cdots\Big)\Big] \nonumber \\
 \end{eqnarray}

 Now we first consider $\varphi^9=\varphi$ which corresponds to a pseudo-scalar meson in the dual gauge theory.
 Since the fluctuation $\varphi^9$ does not mix with any other fields, we could  see the density and asymmetry
 dependence of the meson mass easily. In addition, the in-medium properties of $\varphi$ may reveal some feature of in-medium kaon mass which is related to
 the kaon condensation in dense matter. In case we deal with one light and one intermediate quarks,  $\varphi$ may be considered as kaons in QCD, though
 our model does not possess non-Abelian chiral symmetry. Since the mass of the strange quark is not negligible compared to the intrinsic QCD scale $\Lambda_{QCD}\sim 200$ MeV, the Goldstone boson nature might not be crucial for the mass of kaons.
 Now the relevant part of action is given by
 \begin{equation}
 S_{\varphi^9}=-\mu_6\!\int\!\!d^7\sigma~\textrm{STr}\Big[e^{-\Phi}\!\sqrt{-\textrm{det}(a^{(0)})}\Big(1+\frac{1}{2}\textrm{tr}[(a^{(0)})^{-1}a^{(2)}]
 +\frac{\;T^2}{4}[X^a,X^b]G_{bc}[X^c,X^d]G_{da}\Big)\Big]\;,
 \end{equation}
 where
 \begin{eqnarray}
	a^{(2)}_{rs}&\!=\!&G_{\phi\phi}(\partial_r\varphi+i[\bar{A}_r,\varphi])(\partial_s\varphi+i[\bar{A}_s,\varphi]) \nonumber \\
	&&~-iTG_{yy}G_{\phi\phi}\big\{\partial_r y(\partial_s\varphi+i[\bar{A}_s,\varphi])-(\partial_r\varphi+i[\bar{A}_r,\varphi])\partial_s y\big\}[y,\varphi]\;.
 \end{eqnarray}
 Using the explicit form of metric $a_{rs}$, we evaluate the trace  in the action as
 \begin{eqnarray}
 	\textrm{tr}[(a^{(0)})^{-1}a^{(2)}]&\!=\!&a_{(0)}^{rs}a^{(2)}_{sr}=a_{(0)}^{tt}a^{(2)}_{tt}+a_{(0)}^{t\rho}a^{(2)}_{\rho t}+a_{(0)}^{\rho t}a^{(2)}_{t\rho}+a_{(0)}^{\rho\rho}a^{(2)}_{\rho\rho} \nonumber \\
 	&\!=\!&\!\frac{1}{\;G_{tt}\{G_{\rho\rho}+G_{yy}(y^\prime)^2\}+(T^{-1}A^\prime)^2} \nonumber \\
 	&&\times[\{G_{\rho\rho}+G_{yy}(y^\prime)^2\}G_{\phi\phi}(\bar{D}_t\varphi)^2-2iG_{yy}G_{\phi\phi}A^\prime y^\prime\bar{D}_t\varphi\,[y,\varphi]+G_{tt}G_{\phi\phi}(\varphi^\prime)^2]\;. \nonumber \\
 \end{eqnarray}
 Here we have omitted the terms
 \begin{equation}
	T^{-1}G_{\phi\phi}A^\prime(\varphi^\prime\bar{D}_t\varphi-\bar{D}_t\varphi\,\varphi^\prime)
 \end{equation}
  due to the \textit{symmetrized} trace.
 The remaining part of the action is calculated to be
 \begin{equation}
	[X^a,X^b]G_{bc}[X^c,X^d]G_{da}=-2G_{yy}G_{\varphi\varphi}[y,\varphi]^2\, .
 \end{equation}
 Now we are ready to read off the Lagrangian  up to quadratic order in $\varphi$.
 \begin{eqnarray}	
  \mathcal{L}^{(2)}_{\varphi_\pm}\!\!&\!=\!&\!\!\frac{1}{2}\big[\bar{F}D^{-1}\{G_{\rho\rho}+G_{yy}(y^\prime)^2\}
 G_{\phi\phi}\big]_{y_\pm}(\dot{\varphi_\pm})^2+\frac{1}{2}\big[\bar{F}D^{-1}G_{tt}G_{\phi\phi}\big]_{y_\pm}({\varphi_\pm}^\prime)^2 \nonumber \\
	&\equiv&\!\!\frac{1}{2}\mathbb{P}_\pm(\dot{\varphi_\pm})^2+\frac{1}{2}\mathbb{R}_\pm({\varphi_\pm}^\prime)^2 \\
  \mathcal{L}^{(2)}_{\varphi_{12,21}}\!\!&\!=\!&\!\!\frac{1}{2}\big(\big[\bar{F}D^{-1}\{G_{\rho\rho}+G_{yy}(y^\prime)^2\}G_{\phi\phi}\big]_{y_+}+\big[\bar{F}D^{-1}\{G_{\rho\rho}+G_{yy}(y^\prime)^2\}G_{\phi\phi}\big]_{y_-}\big) \nonumber \\
	 &&~~~~~~~~~~\times\{\dot{\varphi_{12}}\dot{\varphi_{21}}+iA_3(\varphi_{12}\dot{\varphi_{21}}-{\dot\varphi_{12}}\varphi_{21})+(A_3)^2\varphi_{12}\varphi_{21}\} \nonumber \\
	&&-\frac{iy_3}{2}\Big(\big[\bar{F}D^{-1}G_{yy}G_{\phi\phi}A^\prime y^\prime\big]_{y_+}+[\bar{F}D^{-1}G_{yy}G_{\phi\phi}A^\prime y^\prime\big]_{y_-}\Big)(\varphi_{12}{\dot\varphi_{21}}-\dot{\varphi_{12}}\varphi_{21}-2iA_3\varphi_{12}\varphi_{21}) \nonumber \\
	 &&+\frac{1}{2}\Big(\big[\bar{F}D^{-1}G_{tt}G_{\phi\phi}\big]_{y_+}+\big[\bar{F}D^{-1}G_{tt}G_{\phi\phi}\big]_{y_-}\Big){\varphi_{12}}^\prime{\varphi_{21}}^\prime \nonumber \\
	&&+\frac{T^2(y_3)^2}{2}\Big(\big[\bar{F}G_{yy}G_{\phi\phi}\big]_{y_+}+\big[\bar{F}G_{yy}G_{\phi\phi}\big]_{y_-}\Big)\varphi_{12}\varphi_{21} \nonumber \\
	&\equiv&\!\!\frac{1}{2}\mathbb{P}\,\{\dot{\varphi_{12}}\dot{\varphi_{21}}+iA_3(\varphi_{12}\dot{\varphi_{21}}-{\dot\varphi_{12}}\varphi_{21})+(A_3)^2\varphi_{12}\varphi_{21}\} \nonumber \\	 &&-\frac{iy_3}{2}\mathbb{Q}\,(\varphi_{12}{\dot\varphi_{21}}-\dot{\varphi_{12}}\varphi_{21}-2iA_3\varphi_{12}\varphi_{21})+\frac{1}{2}\mathbb{R}\,{\varphi_{12}}^\prime{\varphi_{21}}^\prime+\frac{T^2(y_3)^2}{2}\mathbb{S}\,\varphi_{12}\varphi_{21} \label{fluctLagS}
 \end{eqnarray}
 Here the component fields are defined as
 \begin{eqnarray}
 	\varphi=\varphi_0\tau^0+\varphi_i\tau^i=\frac{1}{2}
 	\left(
		\begin{array}{cc}
			\!\!\varphi_0+\varphi_3\!~&~\!\!\varphi_1-i\varphi_2\!\! \\
			\!\!\varphi_1+i\varphi_2\!~&~\!\!\varphi_0-\varphi_3\!\!
		\end{array}
	\right)\equiv\left(
		\begin{array}{cc}
			\!\!\varphi_+\!~&~\!\!\varphi_{12}\!\! \\
			\!\!\varphi_{21}\!~&~\!\!\varphi_-\!\!
		\end{array}
	\right)\;,
 \end{eqnarray}
 and the common function $\bar{F}$ as
 \begin{equation}
	\bar{F}\equiv e^{-\Phi}\sqrt{-\textrm{det}(a^{(0)})}\;.
 \end{equation}

 Now we turn to gauge field fluctuations.
 \begin{eqnarray}
	\alpha_z(t,\rho)=\left(
		\begin{array}{cc}
			\!\!\alpha_+\!~&~\!\!\alpha_{12}\!\! \\
			\!\!\alpha_{21}\!~&~\!\!\alpha_-\!\!
		\end{array}
	\right)
 \end{eqnarray}
 Like the scalar fluctuation $\varphi^9$, the gauge field fluctuations decouple from the other modes.
 Therefore we can consider this mode only.
 Then non-vanishing $a_{rs}$ in (\ref{pertDet}) are
 \begin{eqnarray}
	a^{(0)}_{rs}&\!=\!&G_{rs}+G_{yy}\partial_r y\partial_s y+T^{-1}(\partial_r\bar{A}_s-\partial_s\bar{A}_r) \nonumber \\
	a^{(1)}_{tz}&\!=\!&T^{-1}\big(\dot{\alpha_z}+i[A,\alpha_z]\big)\equiv T^{-1}\bar{D}_t\alpha_z \nonumber \\
	a^{(1)}_{zt}&\!=\!&-a^{(1)}_{tz} \nonumber \\
	a^{(1)}_{\rho z}&\!=\!&i G_{yy}y^\prime[\alpha_z,y]+T^{-1}\alpha_z^\prime \nonumber \\
	a^{(1)}_{z\rho}&\!=\!&i G_{yy}[\alpha_z,y]y^\prime-T^{-1}\alpha_z^\prime \nonumber \\
	a^{(2)}_{zz}&\!=\!&-G_{yy}[\alpha_z,y]^2\;.
 \end{eqnarray}
 Then quadratic Lagrangian becomes
 \begin{equation}
	 S^{(2)}_{\alpha_z}=-\mu_6\!\int\!\!d^7\sigma~\textrm{STr}\Big[e^{-\Phi}\!\sqrt{-\textrm{det}(a^{(0)})}\Big(-\frac{1}{4}\textrm{tr}[\big((a^{(0)})^{-1}a^{(1)}\big)^2]+\frac{1}{2}\textrm{tr}[(a^{(0)})^{-1}a^{(2)}]\Big)\Big]\;.
 \end{equation}
 Also we can write the trace part explicitly up to overall factor $1/2$
 \begin{eqnarray}
	&&-\big(a_{(0)}^{zz}a^{(1)}_{zt}a_{(0)}^{tt}a^{(1)}_{tz}+a_{(0)}^{zz}a^{(1)}_{z\rho}a_{(0)}^{\rho\rho}a^{(1)}_{\rho z}+a_{(0)}^{zz}a^{(1)}_{zt}a_{(0)}^{t\rho}a^{(1)}_{\rho z}+a_{(0)}^{zz}a^{(1)}_{z\rho}a_{(0)}^{\rho t}a^{(1)}_{tz}\big)+a_{(0)}^{zz}a^{(2)}_{zz} \nonumber \\
	&&~~~=-\big[G^{zz}a^{(1)}_{zt}\;(\frac{G_{\rho\rho}+G_{yy}{y^\prime}^2}{D})\;a^{(1)}_{tz}+G^{zz}a^{(1)}_{z\rho}\;(\frac{G_{tt}}{D})\;a^{(1)}_{\rho z} \nonumber \\
	&&~~~~~~~~~~~~~~~~~~~~+G^{zz}a^{(1)}_{zt}\;(\frac{T^{-1}A^\prime}{D})\;a^{(1)}_{\rho z}+G^{zz}a^{(1)}_{z\rho}\;(-\frac{T^{-1}A^\prime}{D})\;a^{(1)}_{tz}+G^{zz}G_{yy}[\alpha_z,y]^2\big] \nonumber \\
	&&~~~=\frac{1}{T^2 D G_{zz}}\Big[(G_{\rho\rho}+G_{yy}{y^\prime}^2)(\bar{D}_t\alpha_z)^2-2iG_{yy}A^\prime y^\prime\bar{D}_t\alpha_z[y,\alpha_z] \nonumber \\
	&&~~~~~~~~~~~~~~~~~~~~~~~~~~~~~~~~~~~~~~~~~~~~~~-(T^2 G_{tt}G_{\rho\rho}+{A^\prime}^2)G_{yy}[y,\alpha_z]^2+G_{tt}(\alpha_z^\prime)^2\Big]
 \end{eqnarray}
 where $D=G_{tt}(G_{\rho\rho}+G_{yy}{y^\prime}^2)+(T^{-1}A^\prime)^2$.
 Here we have also omitted the terms
 \begin{equation}
	(T^2 D G_{zz})^{-1}\Big[A^\prime(\bar{D}_t\alpha_z\;\alpha_z^\prime-\alpha_z^\prime\;\bar{D}_t\alpha_z)+iG_{tt}G_{yy}y^\prime([y,\alpha_z]\alpha_z^\prime-\alpha_z^\prime[y,\alpha_z])\Big]
 \end{equation}
 under the consideration of the $\textit{symmetrized}$ trace.
 We can read off the Lagrangian for this fluctuation up to quadratic order.
 \begin{eqnarray}
	\mathcal{L}^{(2)}_{\alpha_\pm}&\!=\!&\frac{1}{2}\big[\bar{F}(T^2 DG_{zz})^{-1}\{G_{\rho\rho}+G_{yy}(y^\prime)^2\}\big]_{y_\pm}(\dot{\alpha_\pm})^2+\frac{1}{2}\big[\bar{F}(T^2 DG_{zz})^{-1}G_{tt}\big]_{y_\pm}({\alpha_\pm}^\prime)^2~~~~~~~~~~ \label{fluctLag+-} \\
	\mathcal{L}^{(2)}_{\alpha_{int}}&\!=\!&\frac{1}{2}\big(\big[\bar{F}(T^2 DG_{zz})^{-1}\{G_{\rho\rho}+G_{yy}(y^\prime)^2\}\big]_{y_+}+\big[\bar{F}(T^2 DG_{zz})^{-1}\{G_{\rho\rho}+G_{yy}(y^\prime)^2\}\big]_{y_-}\big) \nonumber \\
	&&~~~~~~~~~~\times\{\dot{\alpha_{12}}\dot{\alpha_{21}}+iA_3(\alpha_{12}\dot{\alpha_{21}}-{\dot\alpha_{12}}\alpha_{21})+(A_3)^2\alpha_{12}\alpha_{21}\} \nonumber \\
	&&-\frac{iy_3}{2}\Big(\big[\bar{F}(T^2 DG_{zz})^{-1}G_{yy}A^\prime y^\prime\big]_{y_+}+[\bar{F}(T^2 DG_{zz})^{-1}G_{yy}A^\prime y^\prime\big]_{y_-}\Big) \nonumber \\
	&&~~~~~~~~~~\times(\alpha_{12}{\dot\alpha_{21}}-\dot{\alpha_{12}}\alpha_{21}-2iA_3\alpha_{12}\alpha_{21}) \nonumber \\
	&&+\frac{1}{2}\Big(\big[\bar{F}(T^2 DG_{zz})^{-1}G_{tt}\big]_{y_+}+\big[\bar{F}(T^2 DG_{zz})^{-1}G_{tt}\big]_{y_-}\Big){\alpha_{12}}^\prime{\alpha_{21}}^\prime \nonumber \\
	&&+\frac{(y_3)^2}{2}\Big(\big[\bar{F}(T^2 DG_{zz})^{-1}(T^2 G_{tt}G_{\rho\rho}+(A^\prime)^2)G_{yy}\big]_{y_+} \nonumber \\
	&&~~~~~~~~~~~~~~~~~~~~~~~~~~~~~~+\big[\bar{F}(T^2 DG_{zz})^{-1}(T^2 G_{tt}G_{\rho\rho}+(A^\prime)^2)G_{yy}\big]_{y_-}\Big)\alpha_{12}\alpha_{21} \nonumber \\
	 &\!=\!&\frac{1}{2}\bar\mathbb{P}\,\{\dot{\alpha_{12}}\dot{\alpha_{21}}+iA_3(\alpha_{12}\dot{\alpha_{21}}-
{\dot\alpha_{12}}\alpha_{21})+(A_3)^2\alpha_{12}\alpha_{21}\} \nonumber \\
	&&-\frac{iy_3}{2}\bar\mathbb{Q}\,(\alpha_{12}{\dot\alpha_{21}}-\dot{\alpha_{12}}\alpha_{21}-2iA_3\alpha_{12}\alpha_{21})+\frac{1}{2}\bar\mathbb{R}\,
{\alpha_{12}}^\prime{\alpha_{21}}^\prime+\frac{(y_3)^2}{2}\bar\mathbb{S}\,\alpha_{12}\alpha_{21} \label{fluctLagV}
 \end{eqnarray}

 \subsection{Meson mass}
 Now we calculate the pseudo-scalar and vector meson masses in asymmetric matter.
 We expect that the off-diagonal meson masses will be sensitive to the asymmetry, which is analogous to $\pi^\pm$ in isospin matter,
 while diagonal part is dependent only on the total number density like $\pi^0$ in isospin matter.
For illustration purpose, we take $\lambda=6$ and $M_{KK}=1$ GeV.
In Appendix \ref{la17A}, we use a bit larger value for the 't Hooft coupling $\lambda=17$.
To discuss pion-like and kaon-like mesons, we consider $m^+/m^-=3$ and $m^+/m^-=30$ in this section.

\begin{figure}[!ht]
\begin{center}
\subfigure[]{\includegraphics[angle=0, width=0.45\textwidth]{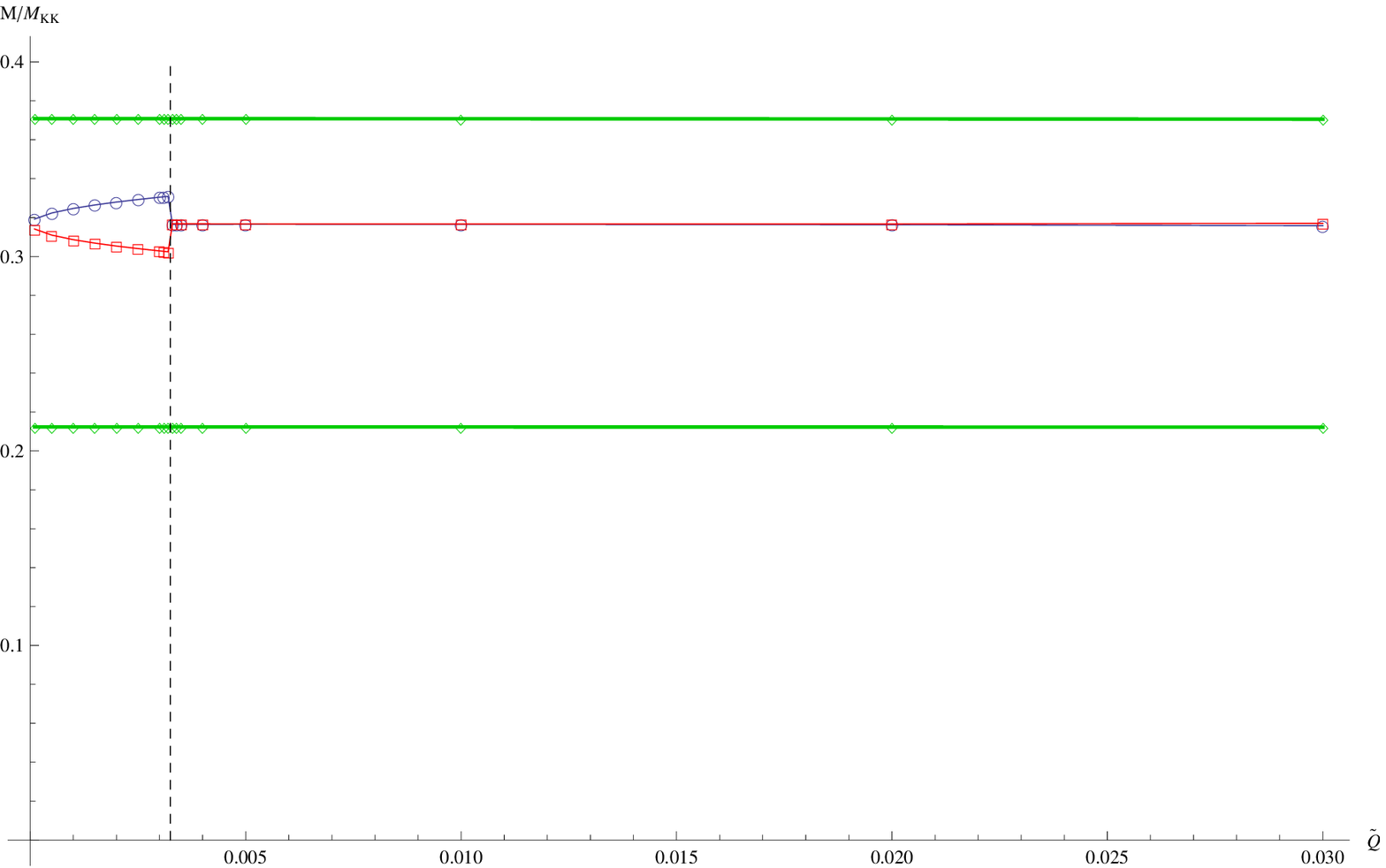}}
~~~~
\subfigure[]{\includegraphics[angle=0, width=0.45\textwidth]{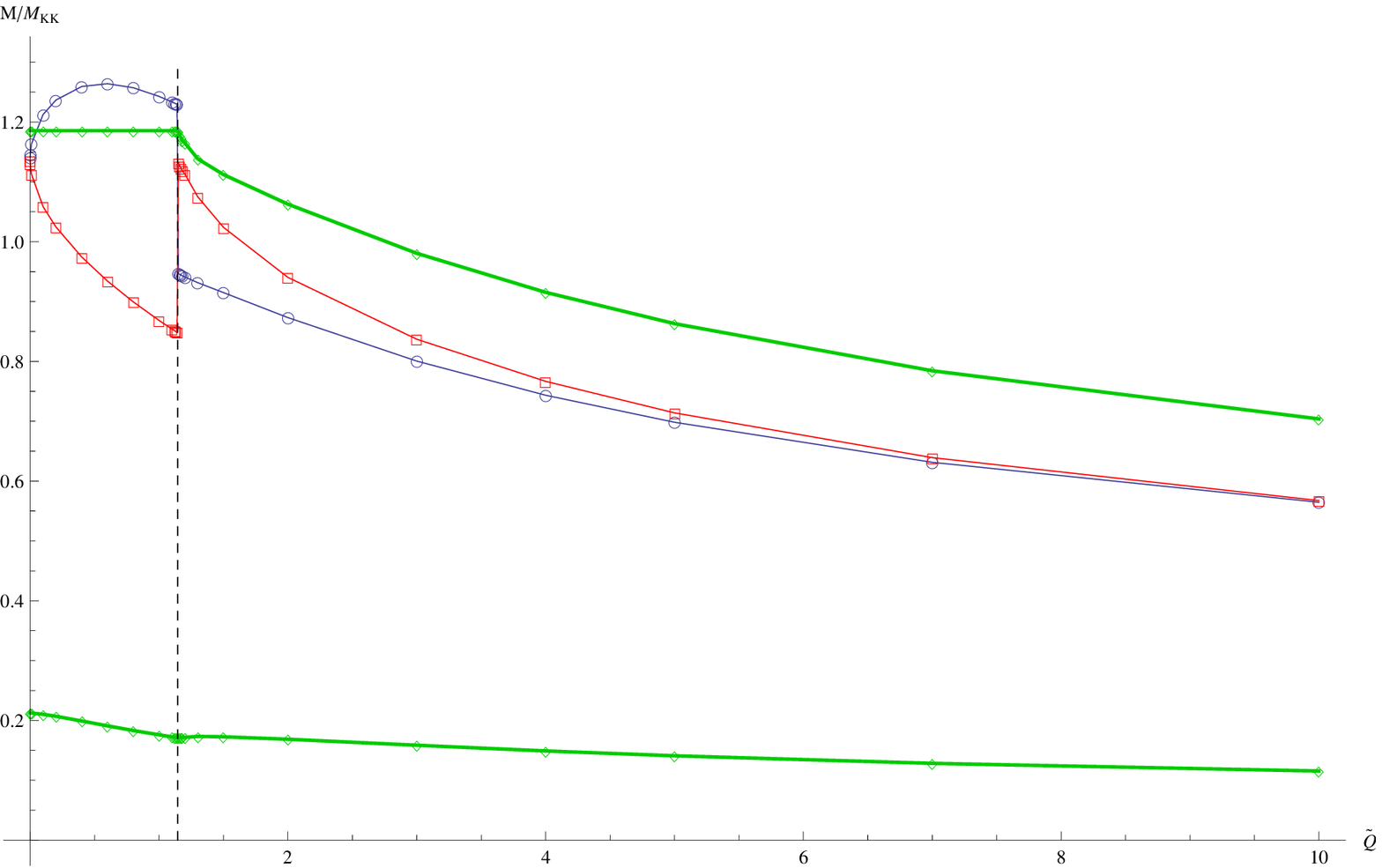}}
\caption{ The in-medium mass of the pseudo-scalar meson with $\lambda=6$ and $M_{KK}=1$ GeV.
Here the vertical dashed line is for the transition density $Q_c$
from $\alpha=0$, maximally asymmetric matter, to $\alpha\neq 0$.
(a) For $m^+/m^-=3$
(b) For $m^+/m^-=30$ \label{la06s}
}
\end{center}
\end{figure}

We first consider the pseudo-scalar meson which is analogous to the pion or kaon depending on the quark mass ratio.
 The equations of motion for $\varphi_{12}$ and $\varphi_{21}$ are, from (\ref{fluctLagS}),
 \begin{eqnarray} \label{eompm}	
  0&\!=\!&\mathbb{P}\,\ddot{\varphi_{12}}+2i(A_3\mathbb{P}-y_3\mathbb{Q})\,\dot{\varphi_{12}}+\mathbb{R}\,\varphi_{12}^{\prime\prime}+\mathbb{R}^\prime\varphi_{12}^\prime+\{-(A_3)^2\mathbb{P}+2y_3A_3\mathbb{Q}-T^2(y_3)^2\mathbb{S}\}\,\varphi_{12}\;, \nonumber \\
  0&\!=\!&\mathbb{P}\,\ddot{\varphi_{21}}-2i(A_3\mathbb{P}-y_3\mathbb{Q})\,\dot{\varphi_{21}}+\mathbb{R}\,\varphi_{21}^{\prime\prime}+\mathbb{R}^\prime\varphi_{21}^\prime+\{-(A_3)^2\mathbb{P}+2y_3A_3\mathbb{Q}-T^2(y_3)^2\mathbb{S}\}\,\varphi_{21}\;.~~~~~~~~
 \end{eqnarray}
 For $\varphi_\pm$, we have
 \begin{equation}
0= \mathbb{P}_\pm\,\ddot{\varphi_\pm}+\mathbb{R}_\pm\,\varphi_\pm^{\prime\prime}+\mathbb{R}_\pm^\prime\varphi_\pm^\prime\;.
 \end{equation}
 If we decompose $\varphi$ as
 \begin{equation}
	\varphi(t,\rho)=e^{-i\omega t}\Phi(\rho)\;,
 \end{equation}
 the equations of motion can be written as
 \begin{eqnarray}	 &&\mathbb{R}\,\Phi_{12}^{\prime\prime}+\mathbb{R}^\prime\Phi_{12}^\prime-[\,(\omega_{12}-A_3)^2\mathbb{P}+2y_3(\omega_{12}-A_3)\mathbb{Q}+T^2(y_3)^2\mathbb{S}\,]\,\Phi_{12}=0\;, \nonumber \\	
 &&\mathbb{R}\,\Phi_{21}^{\prime\prime}+\mathbb{R}^\prime\Phi_{21}^\prime-[\,(\omega_{21}+A_3)^2\mathbb{P}-2y_3(\omega_{21}+A_3)
 \mathbb{Q}+T^2(y_3)^2\mathbb{S}\,]\,\Phi_{21}=0\;,\nonumber\\
 &&  \mathbb{R}_\pm\,\Phi_\pm^{\prime\prime}+\mathbb{R}_\pm^\prime\Phi_\pm^\prime-\omega_\pm^2\mathbb{P}_\pm\Phi_\pm=0\;.
	\end{eqnarray}
As expected the equations for $\varphi_\pm$ are independent of the asymmetry, which is encoded in $A_3$, and depend only on the total charge $Q$.
We solve these equations of motion numerically, and the results are in Fig. \ref{la06s},
where the vertical dashed line is for the transition density $Q_c$
from $\alpha=0$, maximally asymmetric matter, to $\alpha\neq 0$.
Here we choose $M_{KK}=1$ GeV~ and $\lambda=6$ \cite{Kim:2010dp}.
In this case,  $\tilde Q\sim 1.2$ corresponds to the normal nuclear matter density $ \rho_0$.
The relation between the baryon number density and $\tilde Q$ is given by
\begin{eqnarray}
\rho = \frac{2\cdot2^{2/3}}{81 (2\pi)^3} \lambda M_{KK}^3\,\tilde{Q}\, ,\label{rhoQ}
\end{eqnarray}
For illustration purpose, we show low density data only in Fig. \ref{la06s}(a) whose high-density behavior is similar to
Fig. \ref{la06s}(b).
The UV boundary condition is unambiguously fixed by the normalizability condition, $\Phi\sim 1/\rho|_{\rho\rightarrow \infty}$,
 and we choose Neumann boundary condition at IR.

In Fig. \ref{la06s}, green lines are for $\varphi_-$, bound state of two light quarks of the mass $m^-$, and for $\varphi_+$, bound state of two massive quarks
of the mass $m^+$.
Note that the mass of $\varphi_+$ is independent of the density until the transition density. This can be easily understood since the repulsive
force between the U(1) charges of two flavor branes is inactive when $Q<Q_c$.

For the off-diagonal part, both the total charge (or baryon number density) and the asymmetry will affect the meson mass. Here one can regard the asymmetry in $A_3$
as isospin chemical potential or isospin number density. In this case one can easily expect that $A_3$ will induce the mass splitting between $\varphi_{12}$ and $\varphi_{21}$
since they have opposite isospin charges. Analogy is again $\pi^\pm$ with nonzero SU(2) isospin chemical potential in isospin matter:
with increasing $|\mu_I|$, $m_{\pi^+}$ ($m_{\pi^-}$) goes up (down) linearly.
While $A_3$ is causing the mass splitting, the total charge $Q$ would make the mass of $\varphi_{12}$ and $\varphi_{21}$ decreasing as one can see from the behavior
of the diagonal mesons, $\varphi_{\pm}$.
Blue (red) lines in Fig. \ref{la06s} are for $\varphi_{12}$ ($\varphi_{21}$), respectively.
At low densities below $Q_c$, $A_3$ is dominating and the mass splitting is there.
Now as we increase the density $Q$, the asymmetry in $A_3$ is becoming small and so $Q$ is dominating to give almost the same behavior in the mass of both off-diagonal mesons at high densities.

An interesting and/or strange behavior pops up near the transition density. We focus on Fig. \ref{la06s}(b)
since it shows what is going on near the transition density clearly.
The mass of $\varphi_{12}$ suddenly drops and that of $\varphi_{21}$ abruptly goes up, reverting the mass
hierarchy of  $\varphi_{12}$ and $\varphi_{21}$. How can we understand these bizarre results?
Here we try to address this question  based on the Pauli exclusion principle, though
 the behavior near the transition in Fig. \ref{la06s}(b) is too drastic to be explained solely by the Pauli principle.
Since the off-diagonal meson consists of one light (up or down) and one intermediate (strange)  quarks  for $m^+/m^-=30$,
we may consider them as $K^\pm$: $\varphi_{12}$ as $K^+$ ($u \bar s $) and $\varphi_{21}$ as $K^-$ ($s \bar u $).
In Fig. \ref{la06s}(b), the transition happens from pure $up$ quark matter to strange matter with both $up$ and $strange$ quarks.
Based on the Pauli principle between the quarks in the kaons and in the vacuum,
we can expect that after the transition  the mass of $K^-$ would increase since the strange
quarks are piling up in the vacuum, while the mass of $K^+$ drops much faster since the population of the up quarks reduces after the transition.
This can be also seen in a simple model calculation, see Eq. (15) in the first arXiv version of \cite{Kim:2007zzf},
where it is shown that the mass correction
for charged kaons is proportional to $\pm (\rho_u-\rho_s)$.
Here $\rho_q$ is the number density for a quark flavor $q$.

Though the off-diagonal meson in Fig. \ref{la06s}(b) is not exactly the charged kaon in nature partly
because our D4/D6/D6 model is lack of non-Abelian chiral symmetry, we compare our results with
the in-medium charged kaon masses obtained in a QCD effective model.
In the context of an extended relativistic mean field model
the authors of \cite{SM} studied in-medium kaon mass in hyperon-rich dense matter. Their results on $K^\pm$ are qualitatively in agreement with our results
when it comes to a general tendency except near the transition density:
$m_{K^\pm}$ shows mass splitting and then both of them start to decrease at high density.

\begin{figure}[!ht]
\begin{center}
\subfigure[]{\includegraphics[angle=0, width=0.45\textwidth]{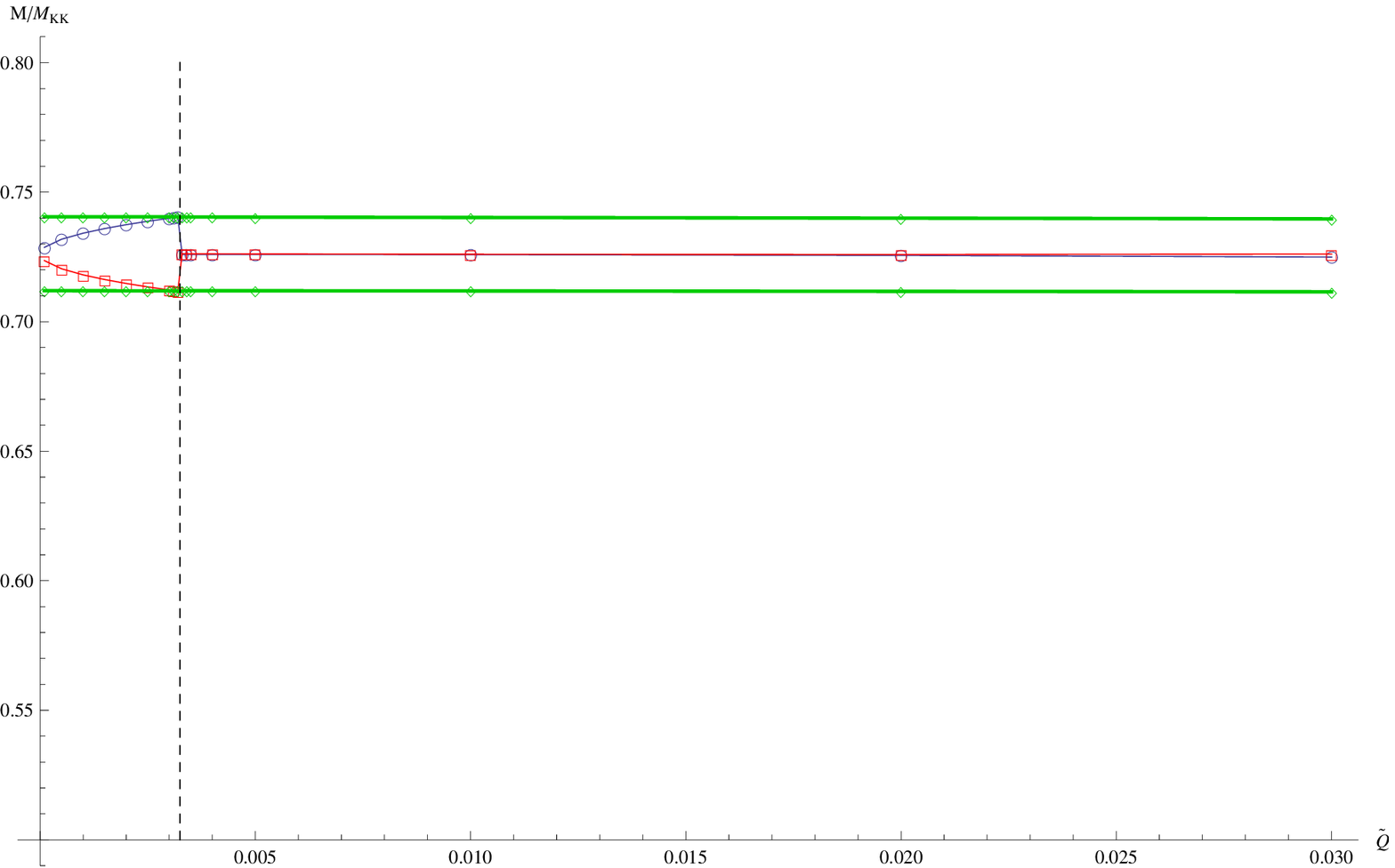}}
~~~~
\subfigure[]{\includegraphics[angle=0, width=0.45\textwidth]{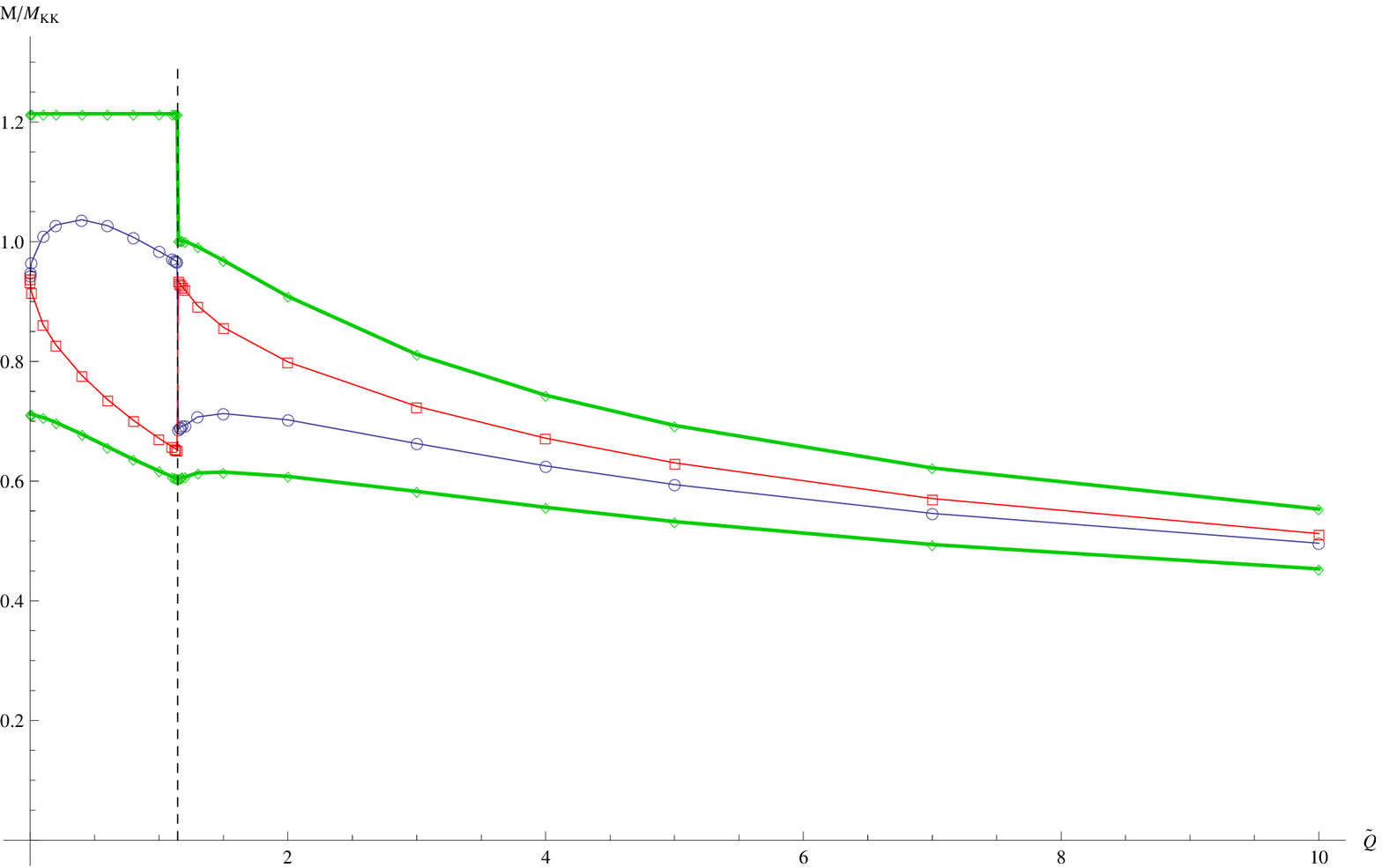}}
\caption{ The in-medium mass of the vector meson with $\lambda=6$ and $M_{KK}=1$ GeV. Here the vertical dashed line is for the transition density $Q_c$
from $\alpha=0$, maximally asymmetric matter, to $\alpha\neq 0$.
(a) For $m^+/m^-=3$
(b) For $m^+/m^-=30$ \label{la06v}
}
\end{center}
\end{figure}

Now we move on to the vector meson. The equations of motion are given by, from (\ref{fluctLag+-}) and (\ref{fluctLagV}),
 \begin{eqnarray}
 	0&=&\bar\mathbb{ R}_\pm\mathcal{A}_\pm^{\prime\prime}
 +\bar\mathbb{R}_\pm^\prime\mathcal{A}_\pm^\prime-\omega_\pm^2\bar\mathbb{P}_\pm\mathcal{A}_\pm\;, \nonumber \\
	 0&=&\bar\mathbb{R}\mathcal{A}_{12}^{\prime\prime}+\bar\mathbb{R}^\prime\mathcal{A}_{12}^\prime-
\{(\omega_{12}-A_3)^2\bar\mathbb{P}+2y_3(\omega_{12}-A_3)\bar\mathbb{Q}+(y_3)^2\bar\mathbb{S}\}\mathcal{A}_{12}\;, \nonumber \\
	 0&=&\bar\mathbb{R}\mathcal{A}_{21}^{\prime\prime}+\bar\mathbb{R}^\prime\mathcal{A}_{21}^\prime-
\{(\omega_{21}+A_3)^2\bar\mathbb{P}-2y_3(\omega_{21}+A_3)\bar\mathbb{Q}+(y_3)^2\bar\mathbb{S}\}\mathcal{A}_{21}\;.
 \end{eqnarray}
Similar to the pseudo-scalar meson, we solve these equations numerically to obtain the result in Fig. \ref{la06v}.
As it is manifest from the figure, the vector meson shows a behavior similar to the pseudo-scalar meson.

\section{Summary}
We studied the in-medium meson mass in a holographic asymmetric dense matter. In this work we considered the pseudo-scalar and
vector mesons.
We used the D4/D6/D6 model with two quark flavors to consider asymmetric dense matter.
A drawback of this model is the absence of the non-Abelian chiral symmetry, while it possesses
the vector $N_f$ flavor symmetry.
We studied a case with $m^+/m^-\sim m_d/m_u$ to see the effect of isospin violation on the meson masses.
Then, we took $m^+/m^-\sim m_s/m_q$, where $m_q\sim(m_u+m_d)/2$, to calculate in-medium kaon-like meson masses.
In both cases we observed the mass splitting of charged pseudo-scalar and vector mesons at low density due to the asymmetry, while
 at high densities charged meson masses become degenerate.
 Near the transition density, we found an exotic behavior in charged meson masses.
 For instance, at the transition density the mass of $\varphi_{12}$ ($K^+$) tumbles down steeply  and that of $\varphi_{21}$ ($K^-$)
 increases very quickly. We could partly understand this behavior based on the Pauli exclusion principle in a boundary quark model picture.

\acknowledgments
Y.K. and I.J.Shin acknowledge the Max Planck Society(MPG), the Korea Ministry of Education, Science, and
Technology(MEST), Gyeongsangbuk-Do and Pohang City for the support of the Independent Junior
Research Group at the Asia Pacific Center for Theoretical Physics(APCTP).
The work of YS and SJS  is supported by the National Research Foundation of Korea(NRF)
grant funded by the Korea government(MEST) (R11-2005-021). 
SJS is also supported by the WCU project (R33-2008-000-10087-0) and   by
NRF  Grant R01-2007-000-10214-0.

\appendix

\section {Mesons with $\lambda=17$ \label{la17A}}
In this Appendix we take  $\lambda=17$ to see if our results are stable with a different value of the t' Hooft coupling constant.
\begin{figure}[!ht]
\begin{center}
\subfigure[]{\includegraphics[angle=0, width=0.45\textwidth]{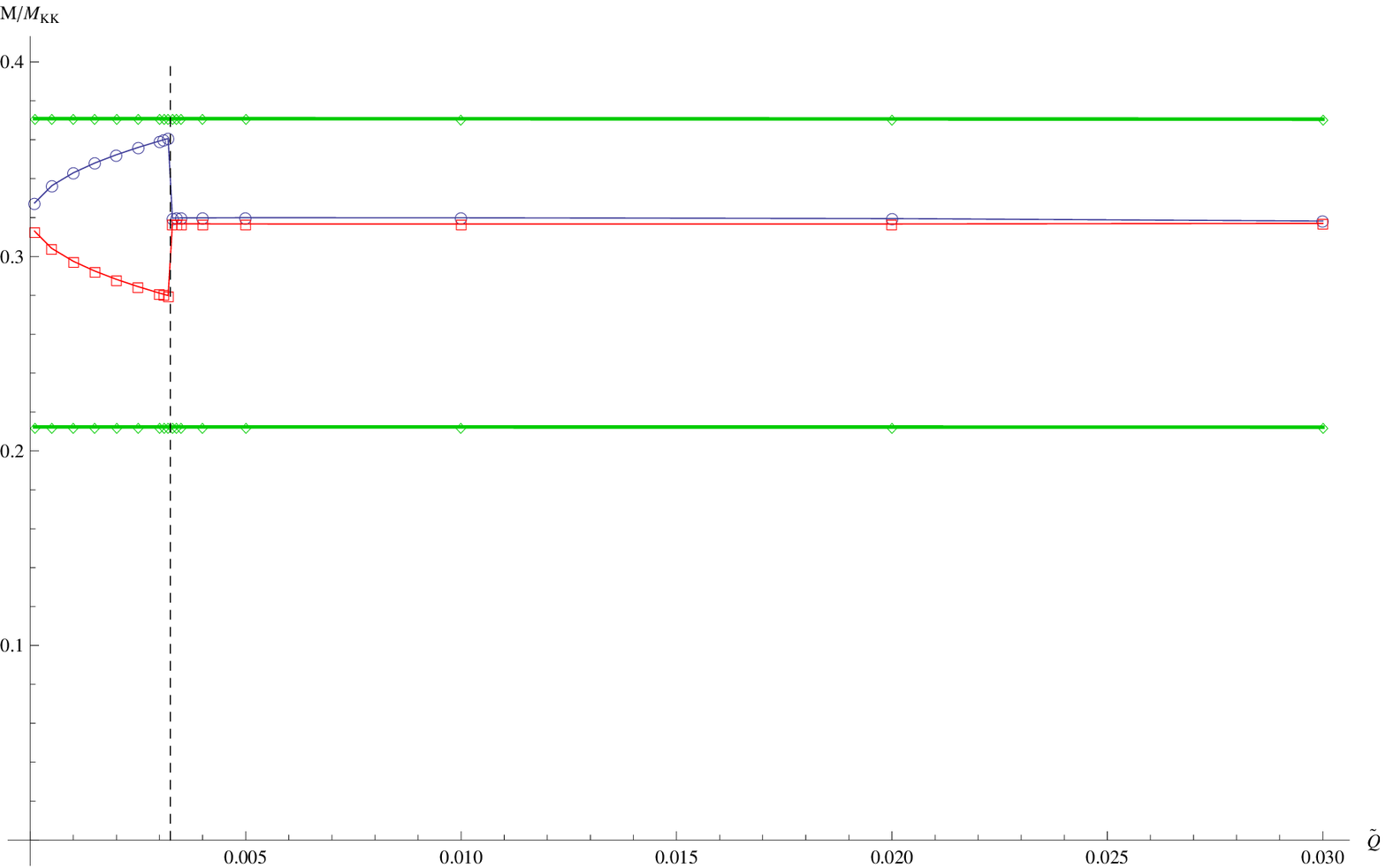}}
~~~~
\subfigure[]{\includegraphics[angle=0, width=0.45\textwidth]{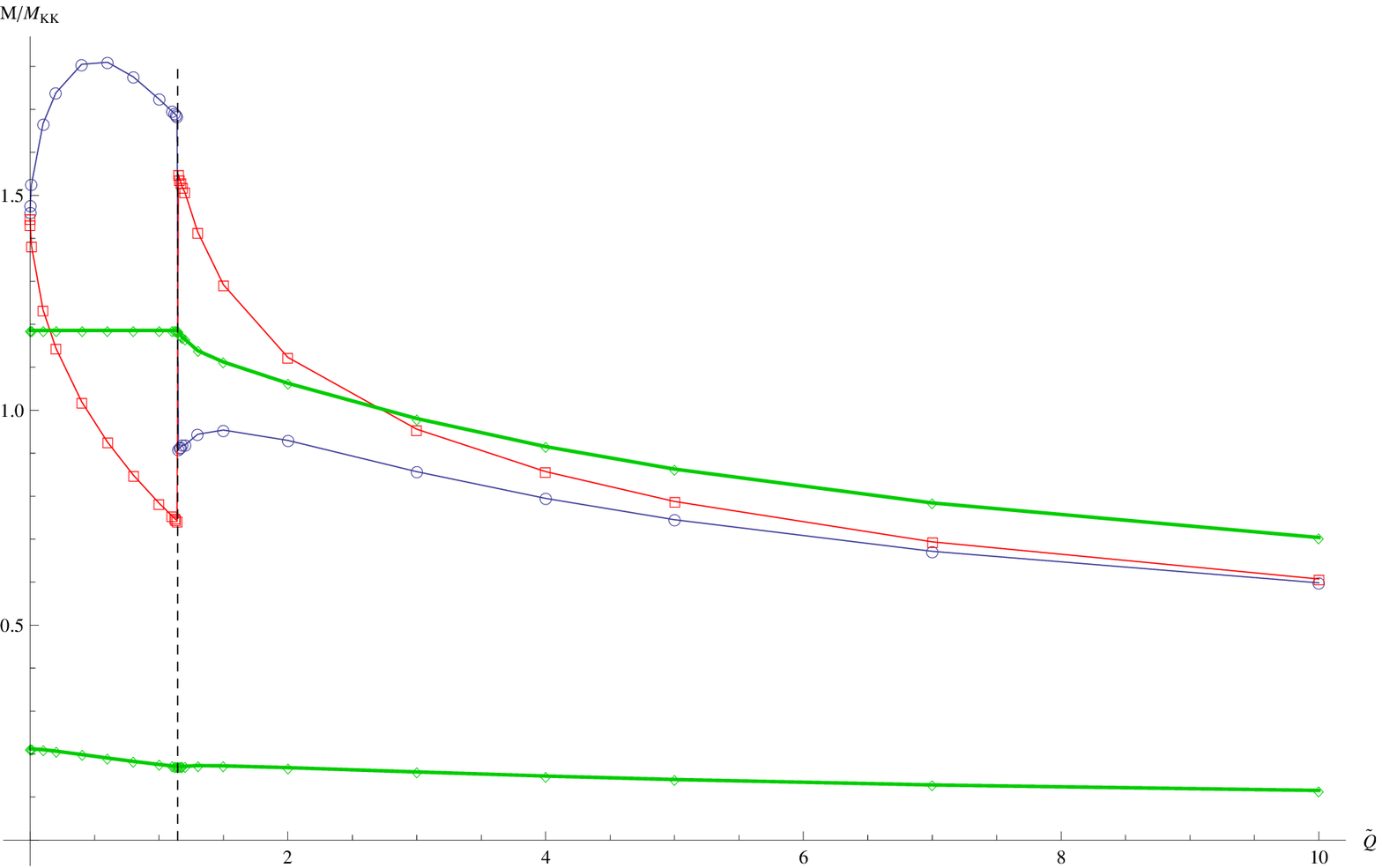}}
\caption{ The in-medium mass of the pseudo-scalar meson with $\lambda=17$ and $M_{KK}=1$ GeV.
Here the vertical dashed line is for the transition density $Q_c$
from $\alpha=0$, maximally asymmetric matter, to $\alpha\neq 0$.
(a) For $m^+/m^-=3$
(b) For $m^+/m^-=30$ \label{la17s}
}
\end{center}
\end{figure}
\begin{figure}[!ht]
\begin{center}
\subfigure[]{\includegraphics[angle=0, width=0.45\textwidth]{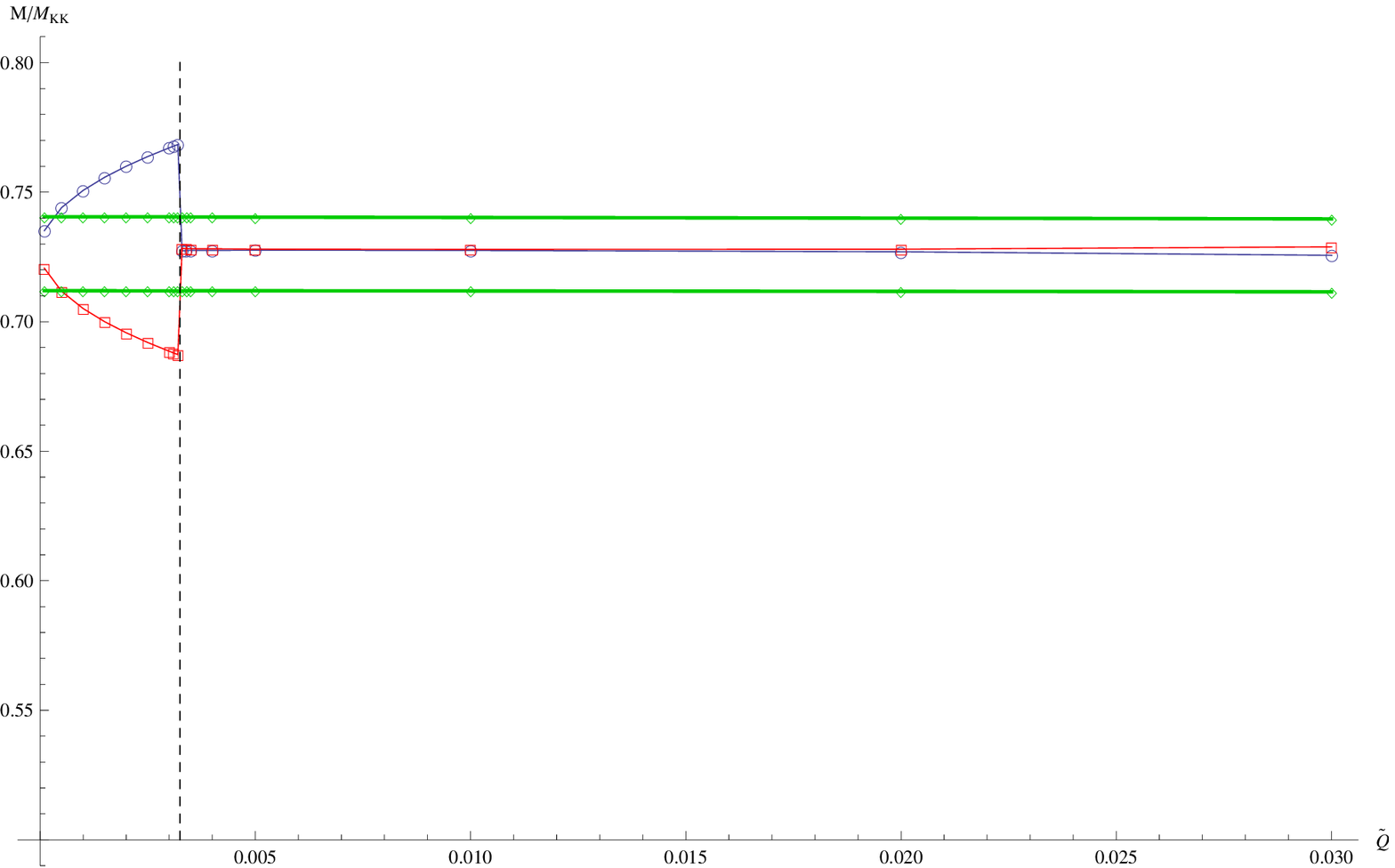}}
~~~~
\subfigure[]{\includegraphics[angle=0, width=0.45\textwidth]{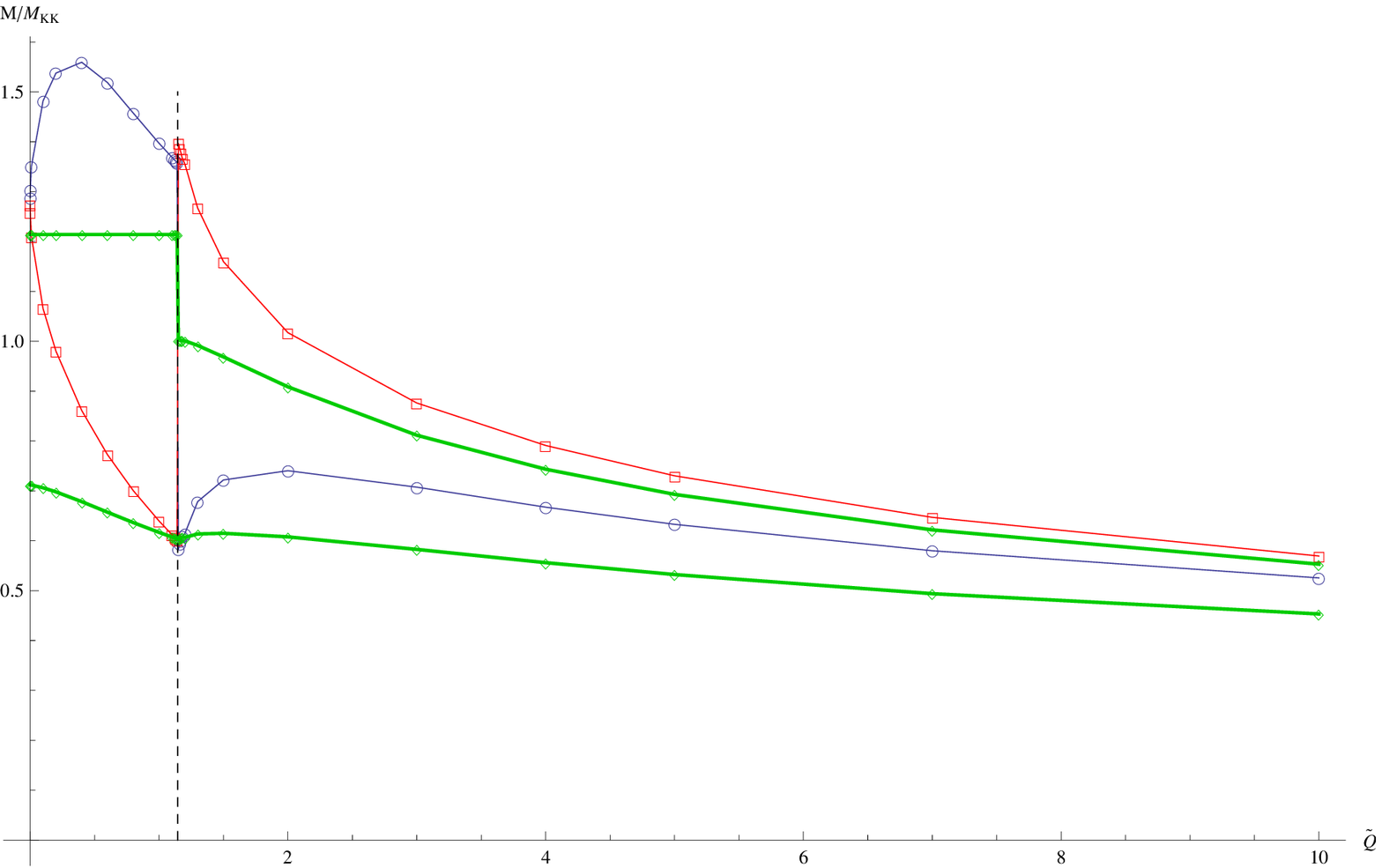}}
\caption{ The in-medium mass of the vector meson with $\lambda=17$ and $M_{KK}=1$ GeV. Here the vertical dashed line is for the transition density $Q_c$
from $\alpha=0$, maximally asymmetric matter, to $\alpha\neq 0$.
(a) For $m^+/m^-=3$
(b) For $m^+/m^-=30$ \label{la17v}
}
\end{center}
\end{figure}

Comparing with the case of $\lambda=6$, we could see that our  results are not that sensitive to a moderate variation of the 't Hooft coupling.

\end{document}